\documentclass[12pt, onecolumn]{article}

\usepackage[utf8]{inputenc}
\usepackage[T1]{fontenc}
\usepackage{graphicx}
\usepackage{dcolumn}
\usepackage{bm}
\usepackage{hyperref}

\usepackage{mathptmx, amsmath, mathrsfs, amssymb,dsfont, mathrsfs}
\usepackage{physics}
\usepackage{braket}
\usepackage{xcolor}
\usepackage{circuitikz}

\usepackage[
backend=bibtex,
style=numeric-comp,
citestyle=numeric-comp,
sorting=none,
giveninits=true,
bibstyle=phys,
biblabel=brackets
]{biblatex}
\AtEveryBibitem{%
  \clearfield{doi}%
  \clearfield{month}%
  \clearfield{issn}%
  \clearfield{isbn}%
}

\DeclareNameAlias{sortname}{family-given}
\DeclareNameAlias{default}{family-given}
\usepackage{float}
\usepackage{epsfig}
\usepackage[utf8]{inputenc}
\usepackage{graphicx,color}
\usepackage{amssymb,amsmath,amsfonts, mathtools, dsfont}
\usepackage[english]{babel}

\usepackage{graphicx}
\usepackage{comment}
\usepackage{braket}
\usepackage{nicematrix,tikz}
\usetikzlibrary{decorations}
\usetikzlibrary{decorations.pathreplacing}
\usepackage{array}
\usepackage{enumitem}

\usepackage{bm}

\usepackage{color}
\usepackage{hyperref}
\hypersetup{
    colorlinks,
    citecolor=red,
    linkcolor=blue,
}

\usepackage[
  a4paper,            
  left=25mm,          
  right=25mm,
  top=25mm,           
  bottom=25mm,
]{geometry}

\usepackage[toc,page]{appendix}

\usepackage{authblk}

\DeclareMathOperator{\Span}{Span}

\newcommand{\mc}{\mathcal}

\newcommand{\p}{\partial}

\newcommand{\C}{\mathbb{C}}
\newcommand{\R}{\mathbb{R}}

\newcommand{\N}{\mathbb{N}}
\newcommand{\ii}{\text{i}}

\begin{document}

\title{Optimal detection of quantum states via projective measurements}

\author[1]{Giuseppe Del Vecchio Del Vecchio}
\author[2]{Satya N. Majumdar}
\vspace{0.2cm}
\affil[1]{%
 Universit\'e Paris-Saclay, CNRS, LPTMS, 91405, Orsay, France
}%

\date{\today}

\maketitle

\begin{abstract}
    We consider the quantum dynamical evolution of a fully-connected quantum system subjected to random projective measurements and study the first detection time of an extended subspace of the Hilbert space. Exact analytical expressions are obtained for the mean first detection time and the full first detection probability distribution as functionals of the initial state $\ket{\psi_0}$ and the measurement rate $r$. Exact enumeration agrees perfectly with our analytical predictions. 
\end{abstract}

\tableofcontents

\section{Introduction}

In recent years, the concept of \emph{stochastic resetting} has emerged as a simple yet powerful protocol that allows to speed up a search process~\cite{evans_2011a, Evans_2011b} and has found various applications across disciplines (see the reviews~\cite{Evans_2020, Pal_2022, Nagar_2023} for an overview). 
The main idea is very simple: the natural time evolution of the system is interrupted at certain time instants $T_1, \dots, T_n$ at which the state of the system is reset to its initial condition (or to a randomly chosen configuration). Usually, these times $T_i$ are chosen such that different periods $\tau_i=T_i-T_{i-1}$ are independent and identically distributed (i.i.d.) random variables with common distribution $f(\tau_i)$. Simple choices are $f(\tau_i)= r\exp(-r\tau_i)$, corresponding to Poissonian resetting \cite{Evans_2020}, and $f(\tau_i) = \delta(\tau_i - T)$ corresponding to `sharp restarts' with period $T$ \cite{Pal_2016, reuveni2016}.
Typically, stochastically resetting the natural evolution of a system has two major consequences. First, the resetting moves manifestly violate detailed balance and drive the system to a non-equilibrium stationary state \cite{Montero2013,Evans2014,Gupta2014,Pal2015,Majumdar2015,Christou2015,Montero2016,Mendez2016,Eule2016,Evans2018,Masoliver2019,Bodrova2019}. 
Second, resetting often expedites the random search of a target. Indeed, one way to measure the performance of a certain search strategy is to study the first time the target is hit~\cite{MAEG09, F2015, GDM2024}. Clearly, this first passage time to the target is a random variable and its probability density function is called first passage probability distribution~\cite{redner2001, metzler2014}. In the simple and common case of Poissonian resetting with rate $r$, the mean first passage time to find the target, as a function $r$, often exhibits a minimum at $r=r^*$ \cite{evans_2011a, Evans_2011b}.
For general stochastic processes, the general conditions for the existence of a non-zero optimal rate $r^*$ have also been elucidated \cite{reuveni2016, besga_2020, Faisant_2021}.
In addition to these theoretical studies, this non-monotonicity of the mean first passage time has also been experimentally demonstrated in a system of correlated particles using optical traps \cite{besga_2020, Tal_Friedman2020, Faisant_2021}.

The reset dynamics can also be studied for quantum systems where the natural time evolution is deterministic and ruled by the Sch\"odinger equation \cite{mukherjee2018}. In this case, the system evolves unitarily but it is stochastically reset to its initial state with rate $r$. As in the classical case, these resets drive the system towards a non-trivial, non-equilibrium stationary state ~\cite{mukherjee2018,rose2018, Dattagupta_2022, Das_2022b, Sevilla_2023}. This basic result  has been applied  in steady-state engineering~\cite{Perfetto2021}, entanglement generation~\cite{kulkarni_2023b} and continuously monitored q-bits~\cite{Dubey_2023}. As mentioned above, stochastic resetting has also the effect of potentially making stochastic searches for targets much more efficient. What about optimizing a quantum search ? Prototypical examples of quantum searches are the Grover's algorithm \cite{grover1996} or a particle on a hypercubic lattice that has to hit the target located at some of the vertices of the hypercube \cite{Giri2017}. Since now the dynamical variable performing the search is a vector in a Hilbert space, i.e., representing the state of the system $\ket{\psi(t)}$, these quantum strategies are very different from their classical counterparts. Nevertheless, they are particularly appealing because, unlike classical random walkers that are diffusive, quantum particles can propagate ballistically so that in principle they are expected to locate targets much faster \cite{kempe}.
This simple observation has motivated the development of quantum random walks~\cite{ambainis_2001, farhi_1998, aharonov_1993} which have also been shown to have exponentially faster hitting times than classical walks~\cite{kempe_2005} and they have been applied to optimize various decision tasks~\cite{farhi_1998, childs_2014}. We address the reader to Refs.~\cite{kempe, vanegas} for comprehensive introductions on quantum random walks. 

\begin{figure}[!h]
    \centering
\includegraphics[scale=1.1]{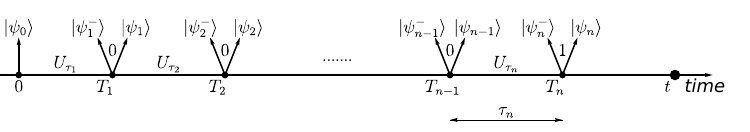}
    \caption{Illustration of the measurement protocol. The target subspace $A$ is probed at times $T_1,\dots, T_n$ up to time $t$ and we stop probing when we successfully detect the state in $A$. The time intervals $\tau_i=T_i - T_{i-1}$ between two measurements are independent with common distribution $f(\tau_i)$ and the last interval $\ell = t - \sum_{i=1}^n \tau_i$ is measurement free. In between two measurements the system evolves unitarily with time evolution given by $U_{\tau_i}=\exp(-\ii \tau_i H)$. The $0$s indicate a failed attempt while the final $1$ indicates the successful detection. The state before the $i$-th measurement is $\ket{\psi^-_i} = \exp(-\ii t H)\ket{\psi_{i-1}}$ while immediately after is $\ket{\psi_i}$ as in Eq. \eqref{eq:projection_measurement}. Each failure happens with probability Eq. \eqref{eq:ith_failure}.}
    \label{fig:measurement_protocol}
\end{figure}

For these quantum walks it is necessary to give a clear definition of what the first passage time is and, as elaborated below, the measurement postulate of quantum mechanics~\cite{sakurai2020, krovi_2006b, Dhar_2015b, Kulkarni_2023} provides a simple answer. Furthermore, the use of measurements makes a special kind of resetting dynamics undertaken by the system appear in a very natural way \footnote{Although other protocols to physically implement a resetting dynamics in a quantum system have been recently proposed \cite{Kulkarni_2025, mesquita2025}.}. This postulate simply tells that if we measure a quantum system and find it in a certain state, then immediately after the measurement the system is projected into that state. 
For concreteness, suppose we have a quantum system with Hamiltonian $H$ in state $\ket{\psi_0}$ at time $t=0$ and the goal is to successfully detect whether the system is in a certain `target' subspace $A$ of the total Hilbert space.  What we mean by successful detection is that the state is found to belong to the subspace $A$. The protocol then consists in probing the system repeatedly at times $T_1, T_2 \dots$ with projective measurements \emph{until} detection. In analogy with stochastic resetting for classical systems, these measurements are separated by random intervals $\tau_i= T_i - T_{i-1}$ each drawn independently from a common distribution $f(\tau_i)$. Many works focused on the stroboscopic protocol corresponding to $f(\tau_i) = \delta(\tau_i - T)$ both theoretically~\cite{krovi_2006, krovi_2006b, Grunbaum2013, Dhar_2015a, Dhar_2015b, Friedman_2017, lahiri_2019, yin2019, dubey_2021, yin_2023, yin_2024} and experimentally~\cite{tornow_2023, wang2024hitting}.
When $f(\tau_i) =r \exp(-r \tau_i)$ the protocol is referred to as Poissonian protocol and will be studied in depth in this work. At each measurement we record the outcome as a $0$ if the detection fails and as a $1$ if the detection is successful. The protocol is stopped when we successfully detect the state in $A$. In this way we obtain a sequence of outcomes $\{0\,,\dots,0,1\}$ and the final $1$ marks the successful detection. More precisely, let the states of the system just before and after a failed measurement performed at time $T_i$ be $\ket{\psi^-_i}$ and $\ket{\psi_i}$ respectively. They are related by the following rules~\cite{Dhar_2015a, Dhar_2015b}
\begin{equation}\label{eq:projection_measurement}
    \ket{\psi^-_i}= \exp(-\ii \tau_i H)\ket{\psi_{i-1}}\quad , \quad \ket{\psi_i} = \frac{(\mathds{1}-P_A)\ket{\psi^-_i}}{\|(\mathds{1}- P_A)\ket{\psi^-_i}\|}
\end{equation}
where $P_A$ is the projector on the target subspace $A$ and the denominator in the above expression is just the square root of the probability of a `failed detection'
\begin{equation}\label{eq:ith_failure}
    p_i^{\rm failure} = \|(\mathds{1}- P_A)\ket{\psi^-_i}\|^2
 = \braket{\psi^-_i|(\mathds{1}- P_A)\psi^-_i} \,.
\end{equation}
The reason behind Eq. \eqref{eq:projection_measurement} is clear: the first equation on the left corresponds to the fact that in between two measurements the system evolves unitarily. The second equation on the right is just the translation of the measurement postulate of quantum mechanics: after a failed attempt, since the state was not found in $A$ by the measurement, it has to be in the orthogonal complement $A^\perp$, i.e., we apply the projector $P_{A^\perp} = \mathds{1}-P_A$ and renormalize in order to have unit norm. What the measurement is doing is effectively resetting the state not to a single fixed state but in a superposition of them living in this complement subspace $A^\perp$. See Fig. \ref{fig:measurement_protocol} for an illustration. In classical resetting systems, the analogous situation to the measurement protocol described above is when at each reset the process is restarted from a certain position $x$, but where $x$ is random with a certain distribution $p(x)$ \cite{evans_2011a, Evans_2011b, Evans_2020, Faisant_2021, besga_2020}, (the standard situation is recovered for $p(x) = \delta(x-x_0)$ where we reset at $x_0$ with certainty). 

To compute the distribution of  first detection time, it turns out  to be convenient to study the survival probability $S(t)$, i.e., the probability that the subspace $A$ stays undetected up to time $t$. For a random measurement protocol and a generic Hamiltonian $H$ described the rules Eq. \eqref{eq:projection_measurement}, it was shown in Ref. \cite{Kulkarni_2023} that $S(t)$ can be expressed as 
\begin{equation}\label{eq:surv_intro}
    S(t) = \sum_{n=0}^\infty\left(\prod_{i=1}^n\int_0^{+\infty}\dd \tau_i\right)  P(n,\{\tau_i\}_{i=1}^n|t)S(n,\{\tau_i\}_{i=1}^n)\,.
\end{equation}
where
\begin{align}\label{eq:surv_cond_intro}
    S(n,\{\tau_i\}_{i=1}^n) = \| \tilde U_{\tau_n}\dots \tilde U_{\tau_1}\ket{\psi_0}\|^2\quad , \quad \tilde U_\tau = P_{A^\perp}U_\tau\quad, \quad U_\tau = \exp(-\ii \tau H)\,,
\end{align}
is  the probability that there are $n$ failed measurements up to time $t$ and that the successive intervals between measurements are given by $\tau_i$. The quantity $P(n,\{\tau_i\}_{i=1}^n|t)$ refers to the probability that there are $n$ measurements in time $t$ separated by intervals $\tau_i$. Note that $S(n,\{\tau_i\}_{i=1}^n)$ is completely deterministic and contains only the information about the quantum evolution of the system. In contrast the quantity $P(n,\{\tau_i\}_{i=1}^n|t)$ is a purely classical object and contains the information about the statistical fluctuations arising from the random measurement protocol. Later, in Section \ref{section:main_concepts}, we provide a derivation of the results in Eq. \eqref{eq:surv_intro} and Eq. \eqref{eq:surv_cond_intro} for the sake of completeness.

Once the survival probability $S(t)$ is known, the probability density $F(t)$ of the first detection time can be obtained as
\begin{equation}\label{eq:F_der_S}
    F(t) = - \p_t S(t)\,.
\end{equation}
This follows from the fact that if the first detection takes place after time $t$, then the survival probability up to time $t$ is given by
\begin{equation}\label{eq:survival_first_intro}
    S(t) = \int_t^{+\infty}F(t') \dd t'\,.
\end{equation}
Consequently, the mean first detection time (MFDT) is given by
\begin{equation}\label{eq:mfdt_intro}
    T = \int_0^{+\infty}\dd t\, t\, F(t) = \int_0^{+\infty}\dd t \, S(t)\,,
\end{equation}
where we used Eq. \eqref{eq:F_der_S} and the fact that $S(0)=1$. For an efficient quantum detection protocol, the goal is to minimize $T$ as a function of the parameters of the random measurement process. 

While the classical part in Eq. \eqref{eq:surv_intro}, $P(n,\{\tau_i\}_{i=1}^n)$ is easy to write explicitly \cite{Kulkarni_2023}, the most challenging task is to compute the quantum evolution part $S(n,\{\tau_i\}_{i=1}^n)$ in Eq. \eqref{eq:surv_cond_intro}. This is because the unitary evolution followed by projective measurements mixes different states, leading to a proliferation of matrix elements that one needs to keep track of as a function of time. For the case of a single qubit, this can be done explicitly as was shown by the authors of Ref. \cite{Kulkarni_2023} and it is useful to briefly summarize their main findings. This will help putting in perspective the new results obtained in this paper. 

\begin{figure}[h]
    \centering
    \includegraphics[width=0.9\linewidth]{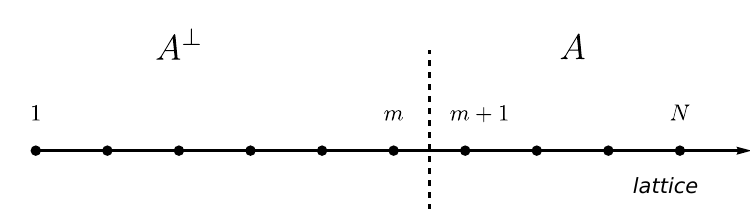}
    \caption{Illustration of the lattice partitioned in the two orthogonal subspaces $A$, associated to sites from $1$ to $m$, and $A^\perp$, associated to sites from $m+1$ to $N$.}
    \label{fig:lattice}
\end{figure}

In Ref. \cite{Kulkarni_2023} a single qubit with a general Hamiltonian was considered. In this case, the Hilbert is described by two states $\ket{\uparrow}$ and $\ket{\downarrow}$. The detector tries to measure if the qubit is the state $\ket{\downarrow}$, i.e., the measured subspace $A=\{\ket{\downarrow}\}$. For a Poissonian measurement protocol with $f(\tau_i) = r\exp(-r \tau_i)$, it was found that the MFDT $T(r)$ exhibits two different behaviors depending on the initial state $\ket{\psi_0}$:
\begin{enumerate}
    \item If $\ket{\psi_0}=\ket{\uparrow}$ then $T(r)$, as a function of $r$, diverges in both limits $r\to 0$ and $r\to + \infty$, with a single minimum at an optimal value $r^*$.
    \item If $\ket{\psi_0}=\ket{\downarrow}$, then $T(r)$ decreases monotonically to zero with increasing $r$. In this case there no finite optimal rate $r^*$.
\end{enumerate}
In addition, the full first detection probability $F(t)$ was also studied and some generic features of $F(t)$ at short and long times were observed. In particular it was found that
\begin{enumerate}
    \item Short times: if $\ket{\psi_0}=\ket{\uparrow}$ then $F(t)\sim t^2$ as $t\to 0$, while if $\ket{\psi_0}=\ket{\downarrow}$ then $F(t)\sim r$ as $t\to 0$.
    \item Long times: $F(t)\sim \exp(-t/t_m(r))$ where the time scale $t_m(r)$, as a function of $r$, has always a unique minimum at a value $r_m^*\neq r^*$, irrespectively of the initial state $\ket{\psi_0}$.
\end{enumerate}
It is then natural to ask how generic are these features of the MFDT $T(r)$ and the first detection probability $F(t)$ when one considers a system larger than a single qubit. For instance, either the Hilbert space may be bigger consisting of multiple basis states or the measured subspace $A$ may contain more than one state. As an example, consider a very simple generalization of the single qubit problem. Suppose we have a single particle on a one dimensional lattice with sites labelled from $1$ to $N$, localized initially at some lattice site. We now place a detector that can measure if the particle is in the sub-lattice consisting of sites from $m+1$ to $N$ with $N\geq m$. In this example, the measured subspace $A$ is just the set of contiguous sites $[m+1,\dots, N]$.
The projection operator onto $A$ is simply $P_A=\sum_{x=m+1}^N \ket{x}\bra{x}$. See Fig. \ref{fig:lattice} for an illustration.
The complementary subspace $A^\perp$, consists of sites $[1,\dots,m]$.
Thus, the measured subspace $A$ consists of $N-m$ states (one for each site of $A$), instead of just one state as in the single qubit problem. Besides, the full Hilbert space consists of $N$ states, going beyond just the two states of the single qubit. We then ask: when is the first time the detector finds the particle if we measure with a Poissonian protocol with rate $r$ ? Even in this seemingly simple problem, computing the first detection probability $F(t)$ turns out to be highly non-trivial, as we will show in this paper, precisely due to the fact that both the Hilbert space and the measured subspace consist of multiple basis states.

In this setting and for a particle with a given Hamiltonian $H$, it is then important to understand the generic features of $F(t)$ (both short and long time behaviors), as well as, the criteria for existence of a unique optimizer $r^*$ of the MFDT $T(r)$.
In this paper we consider a simple all-to-all Hamiltonian and with this choice we are able to solve exactly this model. In particular, we compute the MFDT and the first detection probability distribution $F(t)$. We will see that the criterion for the existence of a finite optimal rate $r^*$, minimizing $T(r)$, requires a non-trivial generalization of the single qubit case, namely the introduction of the so-called \emph{dark} and \emph{bright} states. We will explain the dark and bright states in detail later. It turns out that if the initial state $\ket{\psi_0}$ has any overlap with the these dark states, the MFDT is actually infinite. In order to have a finite MFDT we will henceforth assume throughout this paper that the initial state $\ket{\psi_0}$ has no dark component. Similarly, the short and long time behaviors of $F(t)$ are also affected by the existence of such dark states. Finally, we will see that the dimension $N$ the Hilbert space as well as the dimension $N-m$ of the measured subspace $A$ in this model affect the results non-trivially.

In the next Section we define the model precisely and state the main results. In Section \ref{section:main_concepts} we recall, for the sake of completeness, the derivation of the general result in Eq. \eqref{eq:surv_cond_intro}. Section \ref{section:dark_states} introduces the concept of dark  and bright states. Section \ref{section:mean_field} presents the exact solution for the fully connected model with Poissonian measurements performed on a subset of contiguous sites.  Section \ref{sec:optimal_detection} discusses under which conditions it is possible to optimize the detection process and we calculate exactly the mean first detection time and first detection probability. Section \ref{section:conclusions} offers some future perspectives and concluding remarks.

\section{The Model and the Main Results}\label{sec:main_results}

We have already defined the setup of the model in the Introduction. We consider a one dimensional lattice with sites labelled from $1$ to $N$, see Fig. \ref{fig:lattice}. The measured subspace where the particle can be detected consists of contiguous sites labelled $m+1$ to $N$, i.e., 
\begin{equation}\label{eq:target_subspace}
    A=\cup_{x=m+1}^N\{\ket{x}\} \equiv [m+1,N]\quad , \quad N\geq m\,.
\end{equation}
The complementary subspace $A^\perp$ consists of the first $m$ sites, i.e., 
\begin{equation}\label{eq:perp_subspace}
    A^\perp=\cup_{x=1}^m\{\ket{x}\} \equiv [1,m]\,.
\end{equation}
We consider an all-to-all Hamiltonian
\begin{equation}\label{eq:H_ata}
    H = -J \sum_{x,y=1}^N \ket{x}\bra{y}
\end{equation}
where $J$ sets the energy scale. The unitary evolution under this Hamiltonian allows the particle to hop from any site to any other site. We use the simple Poissonian measurement protocol where the time intervals between two successive measurements $\tau_i$'s are independent and each drawn from the exponential distribution $f(\tau_i)=re^{-r \tau_i}$, where $r$ is the measurement rate. The initial state at time $t=0$ is denoted as $\ket{\psi_0}$ and we assume that it is not a dark state to ensure the MFDT is finite (this point is explained in detail in Section \ref{section:dark_states}).  If the initial state happens to be a dark state the non-detection probability $S(t)$ approaches a non-zero constant as $t\to + \infty$. This means that with a finite non-zero probability an initial dark state remains undetected even at infinite times. Note that for a single qubit model discussed earlier there is no dark state. However, whenever the Hilbert space or the measured subspace is large there might be dark states. Therefore, in order to be detected eventually, the initial state should not have any component in the dark subspace, i.e., it has to be a bright state. See Section \ref{section:dark_states} for a detailed classification of dark and bright states.

\begin{figure}[h]
    \centering
    \includegraphics[scale=0.35]{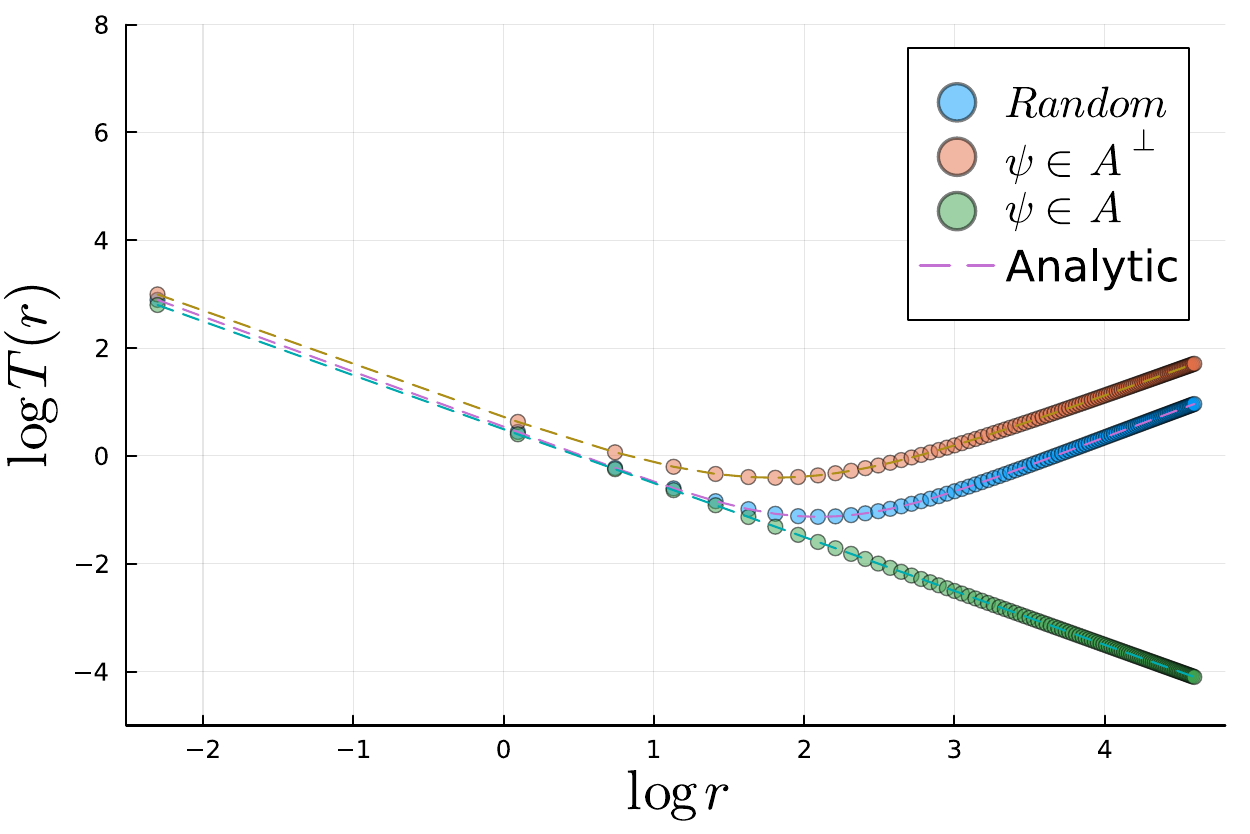}
    \includegraphics[scale=0.35]{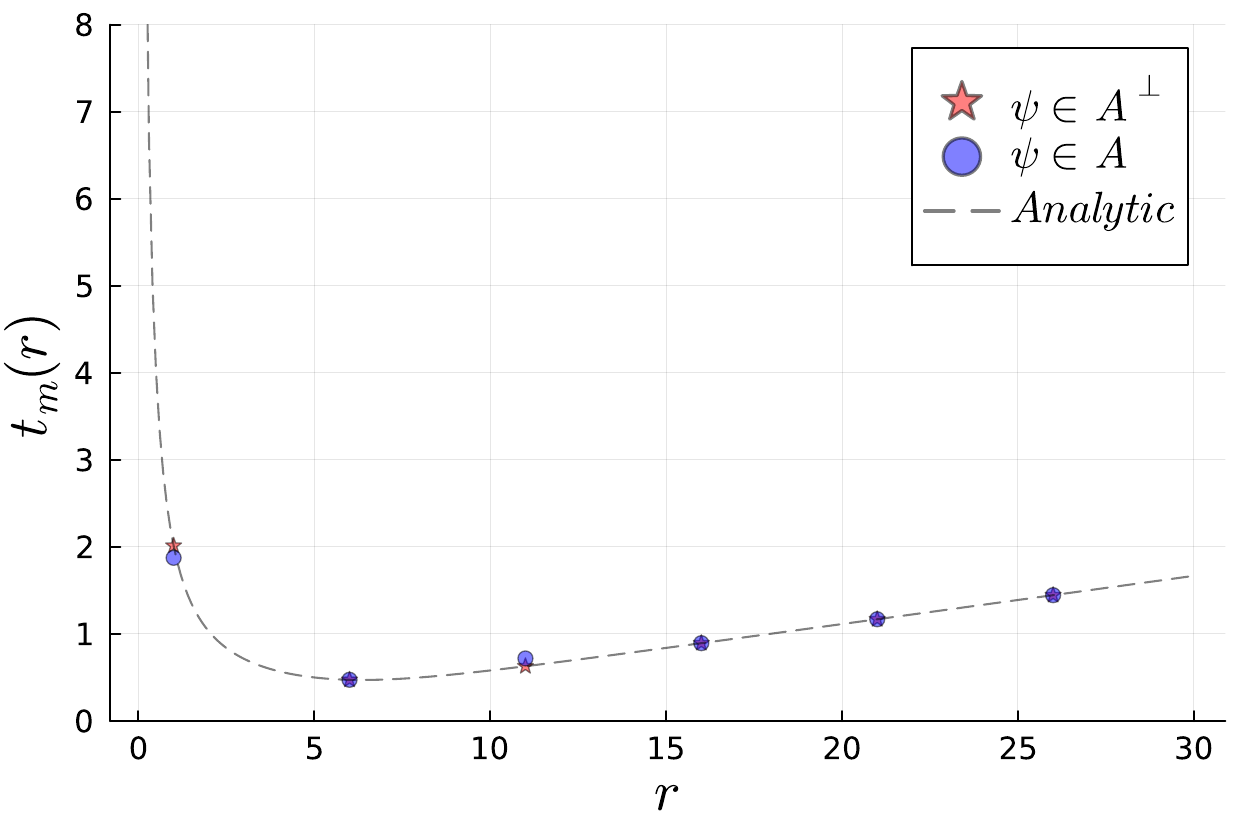}
    \caption{(Left) Analytic prediction Eq. \eqref{eq:mfdt_exact} (dashed) and numerical simulations (points) for the mean first detection time $T(r)$ for different initial bright states localized in $A^\perp$ (red), localized in $A$ (light gree) and a random state (light blue). Logarithms are used to highlight the minimum. (Right) Analytic prediction Eq. \eqref{eq:mfdt_exact} (dashed) and numerical simulations (points) for large time decay time scale of the first detection probability $F(t) \sim \exp(-t/t_m(r))$ in Eq. \eqref{eq:asy_F_large} for different initial bright states, one localized in $A$ one in $A^\perp$. Simulations parameters: $N=6$, $m=3$, $J=1$. }
    \label{fig:mfdt}
\end{figure}

\begin{figure}[h]
    \centering
    \includegraphics[scale=0.35]{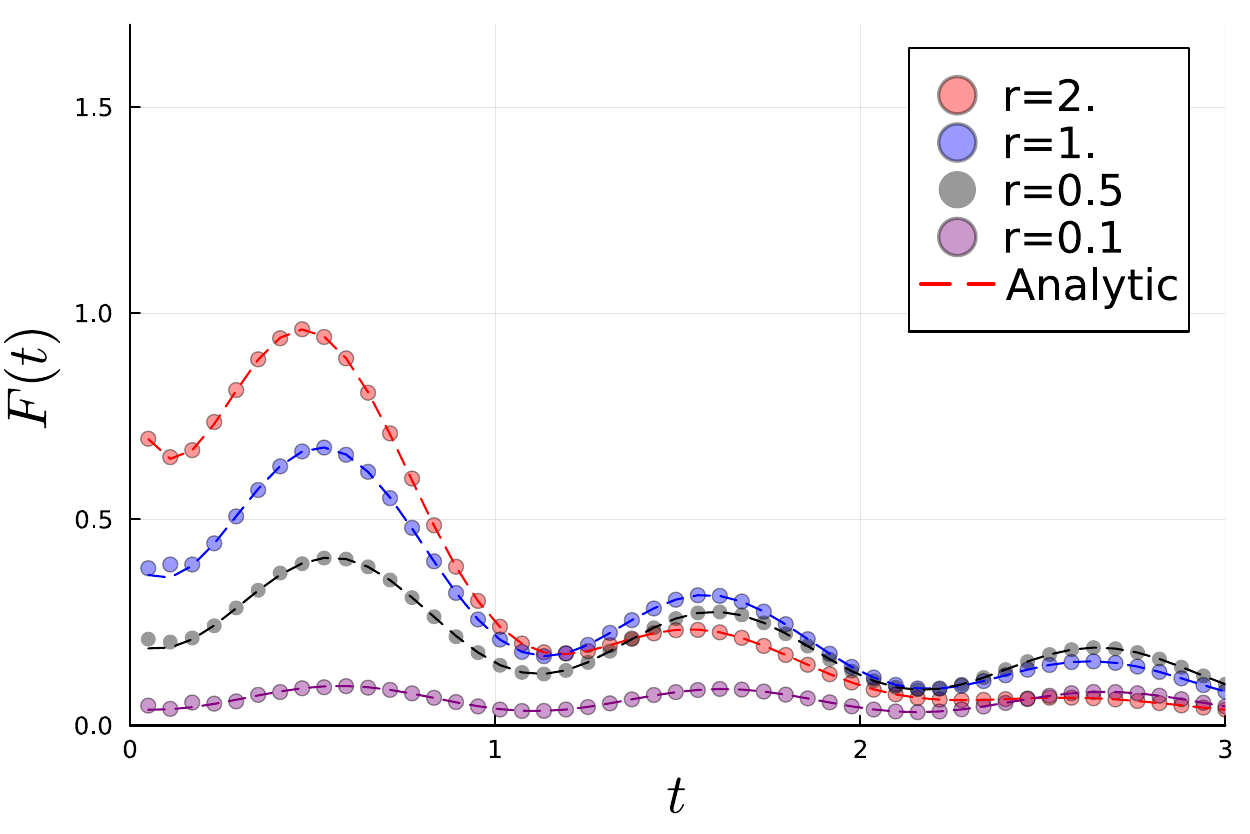}
     \includegraphics[scale=0.35]{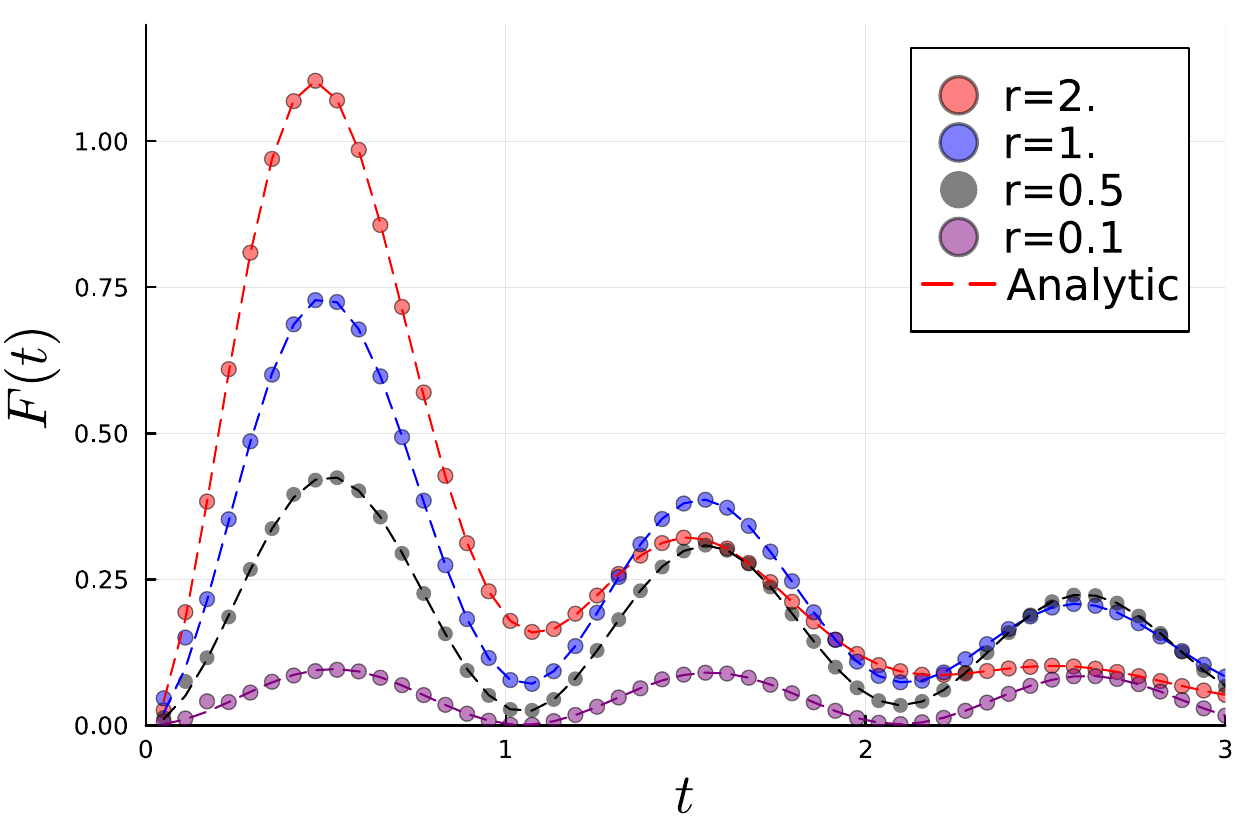}
    \caption{Comparison between analytic prediction (dashed) and numerical simulations (points) for first detection probability $F(t)$ for Poissonian measurements for different rates $r$. The analytical prediction is Eq. \eqref{eq:exact_F}. (Left) Initial state taken as a random bright state. As $t\to 0$, $F(t)\sim \text{const.}$ as predicted by the first line of Eq. \eqref{eq:asy_F_short}. (Right) The special initial state Eq. \eqref{eq:special_state} causing the behavior $F(t) \sim t^2$ as predicted by the second line of Eq. \eqref{eq:asy_F_short}. Simulation parameters are $N=6$, $m=3$ and $J=1$.}
    \label{fig:F_t}
\end{figure}

With this setup and the Hamiltonian in Eq. \eqref{eq:H_ata}, we show that the survival probability $S(t)$ in Eq. \eqref{eq:surv_intro} can be computed exactly for all $t$ and for any initial bright state $\ket{\psi_0}$. From this exact expression, we computed the first detection probability density $F(t) = -\p_t S(t)$ as in Eq. \eqref{eq:F_der_S} and the MFDT $T(r) = \int_0^{+\infty} \dd t\, S(t)$ as in Eq. \eqref{eq:mfdt_intro}. Our main results can be summarized as follows:
\vskip0.3cm
\noindent{\textbf{Mean first detection time (MFDT).}}
The MFDT $T(r)$, as a function of $r$, has the following asymptotic behaviors:
\begin{equation}\label{eq:MFDT_asymptotic_small}
    T(r) \sim \frac{1}{r}\quad , \quad \text{for } r\to 0 \quad\text{and for any initial bright state} \ket{\psi_0}
\end{equation}
where $\sim$ means asymptotic equality up to constants.
In contrast, as $r\to + \infty$ there are two types of asymptotic behaviors depending on the initial bright state $\ket{\psi_0}$:
\begin{equation}\label{eq:MFDT_asymptotic_large}
    T(r) \sim \begin{cases}        r & P_{A^\perp}\ket{\psi_0}\neq 0 \\
       \frac{1}{r} & P_{A^\perp}\ket{\psi_0}=0
    \end{cases}\quad , \quad \text{as } r\to +\infty 
\end{equation}
where $P_{A^\perp}$ is the projection operator on the subspace $A^\perp$ given in Eq. \eqref{eq:perp_subspace}, i.e., $P_{A^\perp} = \sum_{x=1}^{m}\ket{x}\bra{x}$. Thus, when $P_{A^\perp}\ket{\psi_0}\neq 0$, the MFDT diverges in both limits $r\to 0$ and $r\to + \infty$. Indeed, our exact computations show that in this case there is a unique minimum of $T(r)$ at $r=r^*$ given explicitly in Eq. \eqref{eq:rstar}. In contrast, when $P_{A^\perp}\ket{\psi_0}=0$, i.e., the initial state is completely localized in $A$, the MFDT decreases monotonically to zero as $r$ increases (see the left panel of Fig. \ref{fig:mfdt} for both cases). These results, are the direct generalizations of the qubit case discussed in the Introduction.
\vskip0.3cm
\noindent{\textbf{First detection probability density $F(t)$.}}
In analogy with the single qubit system, we computed both the short time and long time behaviors of $F(t)$ for a generic initial bright state $\ket{\psi_0}$. 
\begin{enumerate}
    \item Short times: let us first define a special state $\ket{\psi^*}$ as
    \begin{equation}\label{eq:special_state}
    \ket{\psi^*} = \frac{1}{\sqrt{m}}\begin{pmatrix}
        1 \\ \vdots \\ 1 \\ 0\\ \vdots \\ 0\
    \end{pmatrix}
\end{equation}
which has non-zero components equal to $\frac{1}{\sqrt{m}}$ for the first $m$ sites. Then we find that at short times  
\begin{equation}\label{eq:asy_F_short}
    F(t) \sim  \begin{cases}
        \text{const.}  & \ket{\psi_0}\neq \ket{\psi^*}\\
         t^2 &  \ket{\psi_0}=\ket{\psi^*}
    \end{cases}\quad  \quad \text{as } t\to 0 \,.
\end{equation}
This should be compared to the single qubit case discussed in the Introduction.
\item Long times: in the long time limit we find that $F(t)$ always decays exponentially for any initial bright state $\ket{\psi_0}$
\begin{equation}\label{eq:asy_F_large}
    F(t) \sim e^{ -\frac{t}{t_m(r)}} \quad \text{as }\,\,  t\to +\infty
\end{equation}
where we compute the time scale $t_m(r)$ exactly and find that, as a function of $r$, it always has unique minimum at $r_m^*$, which is different from the value $r^*$ that optimizes
the MFDT, as visible in the right panel of Fig. \ref{fig:mfdt}. In Fig. \ref{fig:F_t} we show $F(t)$ as a function of $t$ for different values of $r$ and different initial states: in the left panel we take random initial states  $\ket{\psi_0}\neq\ket{\psi^*}$ while on the right we take $\ket{\psi_0}=\ket{\psi^*}$ which is completely localized in $A^\perp$. Note how the different short time behavior is different in the two cases as predicted by our calculations. In addition to the short and long time behaviors we also observe that $F(t)$, as a function of $t$, exhibits oscillations which are typical hallmarks of quantum dynamics, see Fig. \ref{fig:F_t}.
\end{enumerate}

\section{First detection probability in a random measurement protocol}\label{section:main_concepts}

In this Section, for the sake of completeness and for setting up our notations, we briefly recall the random projective measurement protocol discussed in Ref. \cite{Kulkarni_2023} and mentioned briefly in the Introduction.

We start with a general quantum system described by a Hamiltonian $H$ and Hilbert space $\mc{H}$. 
The detector tries to detect if the system is in one the states belonging to a certain target subset of the Hilbert space $A\subset \mc{H}$ at random times $T_1, \dots, T_n$. 
The intervals between successive measurements $\tau_i = T_i - T_{i-1}$ are i.i.d. random variables with a common distribution $f(\tau_i)$. We will particularly focus on Poissonian measurement protocol where
\begin{equation}\label{eq:poiss_prot}
    f(\tau_i)=r\, \exp(-r\tau_i)\,.
\end{equation}
The measurement rate $r$ is the only parameter characterizing the statistical nature of the random measurements.
We will denote the orthogonal projector onto the target subspace $A$ by $P_A$ which satisfies
\begin{equation}\label{eq:orth_proj}
    P_A^\dag = P_A\quad ,\quad P_A^2 = P_A\,.
\end{equation}

We start from the system initialized in the state $\ket{\psi_0}$ at time $t=0$. The time evolution for a time $\tau_1$ is given by
\begin{equation}\label{eq:psi_minus}
\ket{\psi_1^-}\equiv\ket{\psi_{\tau_1}} = U_{\tau_1}\ket{\psi_0}\quad ,\quad U_\tau = e^{-\ii \tau H}\,.
\end{equation}
After this time $\tau_1$ we perform a strong projective measurement of the subspace $A$ represented by the projector $P_A$. If we fail to detect the state in $A$, immediately after the measurement the normalized state will be
\begin{equation}\label{eq:state_collapse_1}
     \ket{\psi_1} = \frac{P_{A^{\perp}} \ket{\psi_1^-}}{||P_{A^{\perp}}\ket{\psi_1^-}||}   
 \end{equation}
 where $\|\ket{\psi}\| =\sqrt{\braket{\psi|\psi}}$ and $P_{A^{\perp}} = \mathds{1} - P_A$ is the projection onto the orthogonal complement of $A$ . The effect of the projector is to kill the part of the state that has not been observed. This failure event happens with probability $p_1$ given by
 \begin{equation}
     p_1 = \braket{\psi_1^-| P_{A^{\perp}}|\psi_1^-} = \braket{\psi^+_1| \psi^+_1}  = ||\ket{\psi_1^+}||^2\quad 
 \end{equation}
where in the second equality  we have used $P_A = P_A^2 = P_A^\dag$ and we have defined the non-normalized amplitude
\begin{equation}
    \ket{\psi_1^+} \equiv P_{A^{\perp}}\ket{\psi_1^-}\quad .
\end{equation}
At this point is convenient to define the effective evolution operator is
\begin{equation}\label{eq:effective_time_evolution}
    \tilde U_\tau \equiv P_{A^{\perp}}U_\tau\,.
\end{equation}
From the previous calculation we see that the normalization factor of the state just after the measurement is exactly the square root of the probability of failure $\sqrt{p_1}$. In the second step we do the same, evolve the state for $\tau_2$, do a measurement and condition on failure. The result for the state after the measurement is
 \begin{equation}
     \ket{\psi_2} = \frac{\tilde U_{\tau_2}\ket{\psi_1}}{||\tilde U_{\tau_2}\ket{\psi_1}||}
 \end{equation}
 but this time the probability for this event is obtained multiplying the probability of failure at step 1 times the probability of failure at step 2. This is 
  \begin{align}
     p_2 = p_1 \times\braket{\psi_1|\tilde U_{\tau_2}^\dagger\tilde U_{\tau_2}|\psi_1} &=  p_1 \times \frac{\braket{\psi_1^+|\tilde U_{\tau_2}^\dagger\tilde U_{\tau_2}|\psi_1^+}}{\sqrt{p_1}\sqrt{p_1}}
     \nonumber
     \\
     & = \braket{\psi_2^+|\psi_2^+}
 \end{align}
 where in the second equality we have used Eq. \eqref{eq:state_collapse_1} and in the last line we have defined $\tilde U_{\tau_2}\ket{\psi_1^+}\equiv\ket{\psi_2^+}$. 
Repeating this procedure for $n$ steps gives the probability of non-detection at step $n$
\begin{equation}\label{eq:psi_plus}
    p_n = \braket{\psi_n^+|\psi_n^+}\quad  , \quad \ket{\psi^+_n}\equiv\prod_{i=1}^n \tilde U_{\tau_i} \ket{\psi_0}\quad .
\end{equation}
This quantity clearly depends on the whole history of times $\tau_1,\dots , \tau_n$ and on the final time $t$. Indeed, the number of measurements $n$ up to $t$ is a random variable itself so that it is more appropriate to denote the non-detection probability up to step $n$ as
\begin{align}\label{eq:survival_conditioned}
    S(n,\{\tau_i\}_{i=1}^n) = \|\ket{\psi^+_n}\|^2 = \| \tilde U_{\tau_n}\dots \tilde U_{\tau_1}\ket{\psi_0}\|^2
\end{align}
indicating explicitly that this is the non-detection probability conditioned to having $n$ measurements when we evolve up to a time $t = \sum_{i=1}^n\tau_i + \ell$. Note that Eq. \eqref{eq:survival_conditioned} does not explicitly depend on $t$ since the last time interval $\ell$ is measurement free and the last unitary evolution is given by $U_{\ell}$ which cancels upon taking the scalar product. The non-detection probability up to time $t$ is the average over the joint distribution $P(n, \{\tau_i\}_{i=1}^n|t)$ of the number of clicks and the measurement-free intervals up to time $t$ of Eq. \eqref{eq:survival_conditioned} 
\begin{equation}\label{eq:survival_probability}
    S(t) = \sum_{n=0}^\infty\left(\prod_{i=1}^n\int_0^{+\infty}\dd \tau_i\right)  P(n,\{\tau_i\}_{i=1}^n|t)S(n,\{\tau_i\}_{i=1}^n)\,.
\end{equation}
This provides the derivation of Eq. \eqref{eq:surv_intro} in the Introduction.

To make further progress we need an expression for the joint distribution $P(n,\{\tau_i\}_{i=1}^n|t)$. This can be done as follows. We consider the simplified case of protocol where the time intervals $\tau_i$ are i.i.d. variables with distribution $f(\tau)$. The joint distribution can then be expressed as
\begin{equation}\label{eq:joint_prob_intervals}
    P(n,\{\tau_i\}_{i=1}^n|t)  = f(\tau_1) \dots f(\tau_n) q(\ell) \delta\left(t - \ell - \sum_{i=1}^n \tau_i\right)
\end{equation}
where $q(\ell) = \int_\ell^{+\infty}f(\tau)\dd\tau$ and $\ell$ is the last interval $\ell = t - \sum_{i=1}^n \tau_i$ which is measurement free.
We next insert this explicit form of $P(n,\{\tau_i\}_{i=1}^n|t)$ in Eq. \eqref{eq:survival_probability} and take the Laplace transform of $S(t)$ with respect to $t$
\begin{equation}\label{eq:laplace_transf_def}
    \hat S(s) = \int_0^{+\infty}\dd t \, e^{-st} S(t)\,.
\end{equation}
Using \eqref{eq:survival_conditioned}, we then obtain the simple expression \cite{Kulkarni_2023}
\begin{align}\label{eq:surv_lapla_transf}
    \hat S(s) &= \left[\frac{1-\hat f(s)}{s}\right]\sum_{n=0}^\infty\left(\prod_{i=1}^n\int_0^{+\infty}\dd \tau_i\right)  \prod_{i=1}^n\left[f(\tau_i)e^{-s \tau_i}\right]\| \tilde U_{\tau_n}\dots \tilde U_{\tau_1}\ket{\psi_0}\|^2
\end{align}
where we used $\hat q(s) = \frac{1-\hat f(s)}{s}$, which follows from the definition of $q(\ell) = \int_\ell^{+\infty}f(\tau)\dd\tau$.

Once we have the full Laplace transform $\hat S(s)$ of the survival probability $S(t)$, its moments can be computed by taking derivatives with respect to $s$ and setting $s=0$. For instance, the mean first detection time (MFDT) is given by Eq. \eqref{eq:mfdt_intro}, which is simply $\hat S(0)$. Also, the Laplace transform of the full distribution of the first detection time $F(t)=-\p_t S(t)$ can be expressed in terms of $\hat S(s)$ in Eq. \eqref{eq:surv_lapla_transf}. It then reads
\begin{equation}\label{eq:laplace_first_det}
    \hat F(s) = 1-(1-\hat f(s)) \sum_{n=0}^\infty\left(\prod_{i=1}^n\int_0^{+\infty}\dd \tau_i\right)  \prod_{i=1}^n\left[f(\tau_i)e^{-s \tau_i}\right] \|\tilde U_{\tau_n}\dots \tilde U_{\tau_1}|\ket{\psi_0}\|^2\,.
\end{equation}
This formula simplifies further for the Poissonian measurement protocol where $f(\tau_i) = re^{-r \tau_i}$ and will later be applied for the explicit calculation of the first detection probability for the fully connected Hamiltonian described in the Introduction.

\section{Dark and bright states}\label{section:dark_states}
In this Section we introduce and study  two classes of states that are important for the detection protocol: dark states and bright states. The former are particular states that avoid detection with finite probability. On the other hand, bright states are all states that are orthogonal to dark states \cite{krovi_2006b, krovi_2006, Kessler_2021} and always allow detection with unit probability. 
A short discussion on the relationship between the presence of degeneracy in the spectrum of the Hamiltonian and dark states and a method to find them in such a case is provided in Appendix \ref{app:degeneracy}.
We will shortly see that the dark and bright states are important to understand as they characterize whether the MFDT is finite or not for a given initial state $\ket{\psi_0}$.
\begin{figure}[H]
    \centering
\includegraphics[width=0.7\linewidth]{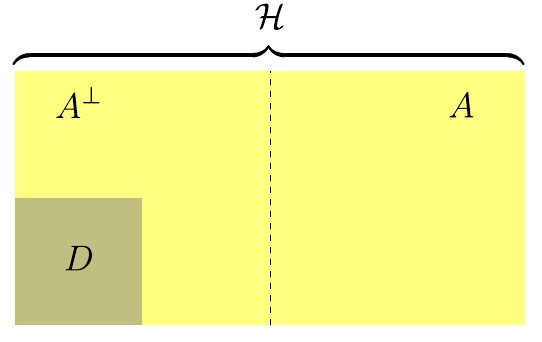}
    \caption{Illustration of the decomposition of the total Hilbert space $\mc{H}$ into the measured subspace and its orthogonal complements in Eq. \eqref{eq:direct_sum_A}. The dark gray area represents states that avoid detection, i.e., the dark subspace in Eq. \eqref{eq:dark_subspace2} while the bright gray area represents its orthogonal complement $B$, the set of bright states, in Eq. \eqref{eq:bright_subspace}.}
    \label{fig:hilbert_space_decomposition}
\end{figure}
For the rest of the discussion we assume that we have a finite dimensional quantum system, with Hilbert space $\mc{H}$ and Hamitlonian $H$. We partition the Hilbert space as
\begin{equation}\label{eq:direct_sum_A}
    \mc{H} = A \oplus A^\perp
\end{equation}
where $A$ denotes the measured subspace where we wish to detect and $A^\perp$ the set of states orthogonal to it and $\oplus$ denotes the \emph{direct sum} of vector spaces.

We now divide the set of eigenvectors of the Hamiltonian $H$ into two different sets:
\begin{itemize}
    \item A dark eigenstate is an eigenstate of the Hamiltonian with \emph{no} component in the measured region $A$.
    \item A bright eigenstate is an eigenstate of the Hamiltonian with \emph{at least} one component in the measured region $A$.
\end{itemize}
Starting from the dark and bright eigenstates we construct the dark and bright subspaces:
\begin{itemize}
    \item The \emph{dark subspace} $D$ is the set of all linear combinations of dark eigenstates. This subspace is represented as a dark grey rectangle in Fig.~\ref{fig:hilbert_space_decomposition}.
    \item The \emph{bright subspace} $B$ is the set of all linear combinations of bright eigenstates.
\end{itemize}

Mathematically the dark subspace takes the following form
\begin{equation}\label{eq:dark_subspace2}
    D \equiv \Span{\{\ket{\mu}\,:\,\, H\ket{\mu} = \lambda \ket{\mu} \,\,,\,\, P_A \ket{\mu} = 0\}}
\end{equation}
where we recall that $P_A$ is the orthogonal projector onto the subspace of measured states $A$.
Since eigenvectors are orthogonal to each other it follows that the bright subspace is the \emph{orthogonal complement} of the dark subspace, i.e.,
\begin{equation}\label{eq:bright_subspace}
    B = D^\perp = \Span\{ \ket{\psi}\in \mc{H}: \braket{\phi|\psi}=0\, ,\, \ket{\phi}\in D\}\,.
\end{equation}
It is useful to consider the projectors onto these subspaces denoted as
\begin{equation}\label{eq:dark_bright_projectors}
    P_D = \sum_{\ket{\mu}\in D}\ket{\mu}\bra{\mu}\quad , \quad P_B = \sum_{\ket{b}\in B}\ket{b}\bra{b}\,.
\end{equation}
From the definitions Eq. \eqref{eq:dark_bright_projectors} one can check that, since different eigenstates are orthogonal, $P_D^2 = P_D$ and $P_D^\dag=P$, i.e., $P_D$ is the \emph{orthogonal} projector operator into the subspace $D$.
Let us list some of the important properties of these two projectors $P_D$ and $P_B$ in Eq. \eqref{eq:dark_bright_projectors}:
\begin{enumerate}[label=(\roman*)]\label{prop:aaa}
    \item $P_D$ commutes with the projector onto the measured subspace, i.e., $[P_A,P_D]=0$. This follows from the definition Eq. \eqref{eq:dark_subspace2}.
    \item $P_D$ commutes with the time evolution operator $U_\tau = e^{-\ii \tau H}$, i.e., $[P_D, U_{\tau_i}]=[P_{B},U_{\tau_i}]=0$. This follows from the fact that $D$ is spanned by dark energy eigenstates.
    \item $P_D$ and $P_B$ projects onto orthogonal subspace, i.e., $P_B P_D=0$. This is clear from  Eq. \eqref{eq:bright_subspace}.
    \item The dark subpace $D$ is a proper subspace of $A^\perp$, i.e., $P_{A^{\perp}} P_{D} = P_{D}P_{A^{\perp}}=P_{D}$. This follows by writing $P_{A^\perp} = \mathds{1} - P_A$ and using the property (i) and from the fact that 
    given two projectors onto two subspaces $C,C'$ (not necessarily orthogonal projectors) then $P_C P_{C'} = P_{C'}$ if and only if $C'$ is a proper subspace of $C$. This is illustrated in Fig. \ref{fig:hilbert_space_decomposition}.
\end{enumerate}

It is important to stress the following fact: a dark state, i.e., a state obtained as a linear combination of dark eigenstates, surely does not overlap with the measured region $A$. On the other hand, given a generic bright state $\ket{\phi}\in B\subset \mc{H}$, it might or might not have components in $A$. This is simply because taking linear combinations of energy eigenstates that \emph{individually} have at least one component in $A$ (and so they are bright eigenstates) it is generically possible to form a state that does not have components in $A$.
This is best illustrated with a simple example: suppose we have a system that only has two bright eigenstates given by
\begin{equation}
    \ket{v_1} = \frac{1}{\sqrt{2}}\begin{pmatrix}
        1 \\0 \\ \vdots\\0  \\ -1
    \end{pmatrix}\quad , \quad \ket{v_1} = \frac{1}{\sqrt{2}}\begin{pmatrix}
        1\\ 0\\\vdots \\0\\ 1
    \end{pmatrix}\,.
\end{equation}
That they are bright means they are annihilated by the projector onto the dark subspace, i.e., $P_D \ket{v_1} = P_D \ket{v_2}=0$. Now suppose the measured subspace $A$ comprises the last coordinate only. Then clearly the linear combination
\begin{equation}
    \ket{\phi}=\frac{1}{\sqrt{2}}\ket{v_1} + \frac{1}{\sqrt{2}}\ket{v_2}
\end{equation}
is still bright because $P_D \ket{\phi}=0$. Nevertheless, we also have $P_A\ket{\phi}=0$ because the last component is zero for $\ket{\phi}$. So this state $\ket{\phi}$ is bright but has no component in $A$.
To test whether a certain state $\ket{\phi}$ is bright one really needs to check if $P_D\ket{\phi}=0$ (or equivalently, to test darkness, $P_B\ket{\phi}=0$).
The different subspaces $A$, $A^\perp$, $D$ and $B$ and their relationship are pictorially represented in Fig.~\ref{fig:hilbert_space_decomposition}. 

The number of basis elements of $D$ gives the dimension $\dim D$ of the dark subspace $D$. Finding this number is very difficult in general because one needs to diagonalize the Hamiltonian and find which of the eigenstates do not overlap with the measured region $A$. Having defined dark and bright states the Hilbert space can also be broken up as
\begin{equation}\label{eq:direct_sum_dark}
    \mc{H}=D\oplus B\,.
\end{equation}
Notice how we have two different ways to write the Hilbert space as a direct sum of orthogonal subspaces, one in terms of the measured subspaces Eq. \eqref{eq:direct_sum_A} and another one in terms of dark and bright states Eq. \eqref{eq:direct_sum_dark} \footnote{Note that the decomposition in Eq. \eqref{eq:direct_sum_A} is independent on the system while \eqref{eq:direct_sum_dark} depends both on $U$ and on the measured subspace $A$.}.

\subsection{Eventual detection and the first detection time}\label{subsec:eventual_detection}

We now illustrate why it is important to understand dark states in the context of the detection protocol. Given an initial state $\ket{\psi_0}$, according to Eq. \eqref{eq:dark_bright_projectors}, we can split it as
\begin{align}\label{eq:splitting_initial_state}
    \ket{\psi_0} = P_D \ket{\psi_0}+P_{B}\ket{\psi_0}
\end{align}
where $P_D$ and $P_B$ are the orthogonal projectors onto the dark and bright subspaces respectively.  To compute the non-detection probability conditioned on $n$ measurements, according to Eq. \eqref{eq:survival_conditioned}, we need to consider the non-normalized amplitude
\begin{equation}
    \ket{\psi_n^+} = \tilde U_n \dots \tilde U_1 \ket{\psi_0}
\end{equation}
where $\tilde U_n$ is the effective non-unitary evolution operator in Eq. \eqref{eq:effective_time_evolution}. Consider first what happens to the part of the state in Eq. \eqref{eq:splitting_initial_state} that lies in the dark subspace, $P_D \ket{\psi_0}$. Recalling the explicit form of $\tilde U_i$ in Eq. \eqref{eq:effective_time_evolution}, we can write
\begin{equation}
    \left\|\tilde U_n\dots \tilde U_1 P_{D}\ket{\psi_0}\right\|^2 =\left\|P_{A^\perp} U_n\dots P_{A^\perp} U_1 P_{D}\ket{\psi_0}\right\|^2\,. 
\end{equation}
By property $(ii)$ of the dark projector $P_D$ (see below Eq. \eqref{eq:dark_bright_projectors})  we can commute $P_D$ with $U_1$ and then by property (iv) we get $P_{A^\perp}P_D = P_D$. Repeating this process of using properties (i) and (iv) up to the right we obtain
\begin{align}
    \left\|\tilde U_n\dots \tilde U_1 P_{D}\ket{\psi_0}\right\|^2 &=\left\| U_n\dots U_1 P_{D}\ket{\psi_0}\right\|^2
    \nonumber
    \\
    & = \|P_D \ket{\psi_0}\|^2
\end{align}
where in the last line we have used the unitary operators preserve the norm. 

As we show below, this implies that for a state that has components in the dark subspace $D$ the mean first detection time is infinite. Indeed, by the previous calculation and
using the decomposition Eq. \eqref{eq:splitting_initial_state} in Eq. \eqref{eq:survival_conditioned} we obtain
\begin{align}\label{eq:sn_scalar_prod}
    S(n, \{\tau_i\}_{i=1}^n) 
    &= \left\| P_{D}\ket{\psi_0}\right\|^2 + \left\| \tilde U_n \dots \tilde U_1\ket{\psi_0}\right\|^2 + 2 \Re \braket{\psi_0|P_D \tilde U_1\dots\tilde U_n \tilde U_n \dots \tilde U_1 P_B|\psi_0}\quad .
\end{align}
Using again properties (i) and (iv), it is easy to see that the third term in Eq. \eqref{eq:sn_scalar_prod} vanishes. This is due to the fact that we can slide $P_D$ up to the right because  $P_D$ and $U_{\tau_i}$ commute. Finally, we also use the fact that $P_D P_B=0$. This then gives
\begin{align}\label{eq:discrete_steps_survival_probability}
    S(n, \{\tau_i\}_{i=1}^n) = \braket{\psi_n^+|\psi_n^+} &= \left\|P_D\ket{\psi_0}\right\|^2 + \left\| \tilde U_n \dots \tilde U_1 P_B\ket{\psi_0}\right\|^2 
\end{align}
which is valid for arbitrary initial states and arbitrary non-unitary $\tilde U_i$. This result just shows that the part of the state lying in the dark subspace is invariant under the time evolution.

An immediate consequence of the above is that if the initial state lies completely in the dark subspace $P_D\ket{\psi_0}=\ket{\psi_0}$, the survival probability is identically $1$ and the mean first detection time is infinite. But the result is stronger: using the expression Eq. \eqref{eq:surv_lapla_transf} for the Laplace transform of the non-detection probability and noting that the first term of Eq. \eqref{eq:discrete_steps_survival_probability} does not depend on the measurement times $\tau_i$ we obtain
\begin{equation}\label{eq:survival_infinite}
    \hat S(s) = \frac{\left\|P_D\ket{\psi_0}\right\|^2}{s} +  \left[\frac{1-\hat f(s)}{s}\right]\sum_{n=0}^\infty\left(\prod_{i=1}^n\int_0^{+\infty}\dd \tau_i\right)  \prod_{i=1}^n\left[f(\tau_i)e^{-s \tau_i}\right]\left\| \tilde U_n \dots \tilde U_1 P_B\ket{\psi_0}\right\|^2  \quad .
\end{equation}
Note that $\hat f(0)=1$ due to the normalization of $f(\tau)$. Assuming that the first moment $\braket{\tau}$ of  $f(\tau)$ exists and is finite, the small $s$ behavior of $\hat f(s)$ is given by
\begin{equation}
    \hat f(s) = 1 - \braket{\tau}s + O(s^2) \quad ,\quad s\to 0\,.
\end{equation}
Consequently, in the $s\to0$ limit,
 $s\hat S(s)$ in Eq. \eqref{eq:survival_infinite} behaves as
\begin{equation}
    \hat S(s) \xrightarrow{s\to0} \frac{\|P_D\ket{\psi_0}\|^2}{s}\,.
\end{equation}
This implies that
\begin{equation}
    S(t) \xrightarrow{t\to+\infty}\|P_D\ket{\psi_0}\|^2 \leq 1\,.
\end{equation}
Consequently, the MFDT in Eq. \eqref{eq:mfdt_intro} diverges. This means that if $\|P_D\ket{\psi_0}\|^2$ is non-zero the MFDT is infinite. Hence, to obtain a finite MFDT, one must have $P_D \ket{\psi_0}=0$, i.e., the initial state should not have any dark component.

A non-zero $\|P_D\ket{\psi_0}\|^2$ also means that there is a finite probability the target stays undetected as $t\to +\infty$. Indeed, the eventual detection probability as $t\to + \infty$ is given by
\begin{equation}\label{eq:eventual_det_dark}
    P_{\rm det}= 1- S(+\infty)= 1 - \|P_D\ket{\psi_0}\|^2\,.
\end{equation}
This result means that we eventually detect target with probability $1$  if an only if $P_D \ket{\psi_0}=0$, i.e., the initial state has no overlap with the dark subsapce $D$. This is in agreement with the conclusions of Refs. \cite{krovi_2006b, Thiel_2020}.

Let us remark that in Ref. \cite{Grunbaum2013} the authors studied the `recurrence problem' for a stroboscopic protocol, i.e., $f(\tau) = \delta(\tau-T)$, where the system is initialized in the state $\ket{\psi_0}$ and the detection subspace is taken to be equal to the initial state, i.e., the projection onto the target subspace is $P_A=\ket{\psi_0}\bra{\psi_0}$. There,  the quantum system was called recurrent if the eventual detection probability $P_{\rm det}=1$ and transient otherwise. It was concluded that for a finite system, for this recurrence problem, $P_{\rm det}=1$ always. We now show that this conclusion also follows immediately from the general analysis presented above. 
Indeed, for $P_A = \ket{\psi_0}\bra{\psi_0}$ one finds, from Eq. \eqref{eq:effective_time_evolution}
\begin{equation}
    \tilde U_{\tau_i} \ket{\psi_0} = P_A U_{\tau_i} \ket{\psi_0} = \ket{\psi_0}\bra{\psi_0} \sum_{\lambda}\lambda_{\tau_i}\ket{\lambda}\braket{\lambda|\psi_0} = \ket{\psi_0}\sum_\lambda \lambda_{\tau_i}|\braket{\psi_0|\lambda}|^2
\end{equation}
where $\{\ket{\lambda}\}$ are the eigenstates of the unitary time evolution $U_\tau$ with eigenvalues $\lambda_\tau$.
Iterating this procedure for all time steps and plugging in Eq. \eqref{eq:surv_lapla_transf} gives
\begin{equation}\label{eq:surv_recurrence}
    \hat S(s) =\left[\frac{1-\hat f(s)}{s}\right]\sum_{n=0}^\infty\left(\prod_{i=1}^n\int_0^{+\infty}\dd \tau_i\right)  \prod_{i=1}^n\left[f(\tau_i)e^{-s \tau_i}\right]\left|\prod_{i=1}^n\sum_\lambda\lambda_{\tau_i}\braket{\psi_0|\lambda}\right|^2\,.
\end{equation}
For this stroboscopic protocol, $\hat f(s) = e^{-s T}\approx 1 - T\, s$ as $s\to 0$. Consequently, from Eq. \eqref{eq:surv_recurrence}, one sees that 
\begin{equation}
    \hat S(0) =  \int_0^{+\infty} S(t) \dd t = T \sum_{n=0}^{+\infty}\left|\prod_{i=1}^n\sum_\lambda\lambda_{T}\braket{\psi_0|\lambda}\right|^2
\end{equation}
is finite and hence the MFDT is also finite. The fact that $\int_0^{+\infty}S(t) \dd t$ is finite indicates that $S(t)\to 0$ faster then $\frac{1}{t}$ as $t\to + \infty$. Hence, $S(+\infty)=0$ and from Eq. \eqref{eq:eventual_det_dark} we see that the eventual detection probability $P_{\det}=1$.

Thus, the main conclusion of this subsection is that a target subspace $A$ will be definitely detected if and only if the initial state has zero overlap with the dark subspace, i.e., $P_D\ket{\psi_0}=0$, and the MFDT will also be finite in that case.

\section{Exactly solvable fully connected model}\label{section:mean_field}
In this Section we go back to our original problem of studying the Poissonian measurement protocol applied to the all-to-all Hamiltonian in Eq. \eqref{eq:H_ata} and extended target subspace $A$ in Eq. \eqref{eq:target_subspace}. In the previous Section, we have seen that in order to have a finite MFDT to detect the target subspace $A$, the initial state has to be bright. For this reason, for the rest of this Section, we will take the initial state $\ket{\psi_0}$ to be a bright state of the Hamiltonian. 
Our main goal is to derive an exact and explicit expression for the Laplace transform $\hat S(s)$ of the non-detection probability in Eq. \eqref{eq:surv_lapla_transf}. From this we can derive the MFDT and the first detection probability as explained in Section \ref{section:main_concepts}. The expression for $\hat S(s)$ in Eq. \eqref{eq:surv_lapla_transf} involves the term $\|\tilde U_{\tau_n}\dots \tilde U_{\tau_1}\ket{\psi_0}\|^2$ where $\tilde U_{\tau}$ is defined in Eq. \eqref{eq:effective_time_evolution} and its calculation constitutes the main technical challenge that we will solve in this Section.

To start, it is convenient to rewrite the
Hamiltonian in Eq. \eqref{eq:H_ata} as
\begin{equation}\label{eq:H_ata_E}
    H = - J \sum_{x,y=1}^N\ket{x}\bra{y} = - J E
\end{equation}
where we have introduced the $N\times N$ matrix $E = \sum_{xy,=1}^N \ket{x}\bra{y}$, i.e., the rank-$1$ matrix with all $1$. 
Expanding the exponential in Taylor series and using $E^{m}=N^{m-1}E$ for any $m\in \N$, one can easily show that
\begin{equation}\label{eq:unitary_evolution_mean_field}
    U_t = e^{\ii J t E} = \mathds{1}+ b_t E
\end{equation}
where $\mathds{1}$ is the $N\times N$ identity matrix
and
\begin{equation}\label{eq:definition_b_mean_field}
    b_t = \frac{e^{\ii Jt N}-1}{N}\,.
\end{equation}
These two observations allow to write the effective time evolution operator $\tilde U_\tau = P_{A^\perp}U_\tau$ in Eq. \eqref{eq:effective_time_evolution} as
\begin{equation}\label{eq:effective_evolution_operator}
    \Tilde{U}_t = P_{A^{\perp}} U_t = P_{A^{\perp}} + b_t P_{A^{\perp}}E
\end{equation}
where we recall that $P_{A^{\perp}}$ is the projector onto the un-measured sub-lattice $ A^{\perp}=\cup_{x=1}^m\{\ket{x}\}\equiv [1,m]$ (see also Fig. \ref{fig:lattice}). 

Let $\{\ket{b}\}$ be a basis of bright states $B = D^\perp$ for the Hamiltonian in Eq. \eqref{eq:H_ata_E}. A generic initial bright state can be written as
\begin{equation}\label{eq:bright_state_expansion}
    \ket{\psi_0} = \sum_{\ket{b}\in B}\ket{b}\braket{b|\psi_0}\,.
\end{equation}
Since we have to calculate $\|\tilde U_{\tau_n}\dots \tilde U_{\tau_1}\ket{\psi_0}\|^2$, we can first compute $\ket{\psi^+_n}= \tilde U_{\tau_n}\dots \tilde U_{\tau_1}\ket{\psi_0}$ (see Eq. \eqref{eq:psi_plus}) and then extract the norm squared. To do this in an efficient manner, we can exploit the linearity of the effective time evolution operator $\tilde U_{\tau}$: we can calculate its repeated action on each basis state $\ket{b}$ in Eq. \eqref{eq:bright_state_expansion} independently and sum these individual results at the end. Before doing this, we need to find a basis of the bright subspace $B$ for fully connected Hamiltonian Eq. \eqref{eq:H_ata}.
As explained in Section \ref{section:dark_states}, a basis for the bright subspace is the set of all eigenstates with at least one non-zero component in the measured region $A$, i.e., in our problem the sub-lattice $[m+1,N]$. The eigenstates and eigenvalues of the Hamiltonian $H$ in Eq. \eqref{eq:H_ata} are calculated in Appendix \ref{appendix:nice_matrix}. We find that the eigenstates are
\begin{equation}
    \ket{\Lambda,l}\quad , \quad l=1,\dots, g_{\Lambda}
\end{equation}
where $\Lambda=0,N$ are the only two eigenvalues while $g_{\Lambda}$ is the degeneracy. The eigenspace associated to $\Lambda=N$ has $g_N=1$ and we denote its unique eigenvector as $\ket{\Lambda=N}$ and it is given by 
\begin{equation}
    \ket{N} = \frac{1}{\sqrt{N}}\begin{pmatrix}
        1 \\ \vdots \\ 1\
    \end{pmatrix}\,.
\end{equation} 
This is simply the vector with identical compontents over all sites and normalized with unit norm.
The remaining states $\ket{0,l}$ have zero eigenvalue and degeneracy $g_0=N-1$. They are given in Eq. \eqref{eq:orthogonalised_overlap} which can be compactly written as
\begin{equation}\label{eq:normalized_degenerate_states}
    \ket{0,l}=C_l \begin{pmatrix}
        -1\\ \vdots\\-1 \\ l \\ 0 \\ \vdots\\0 
    \end{pmatrix}\quad , \quad C_l=\frac{1}{\sqrt{l(l+1)}}\quad, \quad l=1,\dots, N-1
\end{equation}
where in the column vector above the first $l$ components are $-1$ and the $l+1$-th is equal to $l$ and the remaining $N-l-1$ are equal to $0$.

From the explicit expression of the eigenstates, it is then easy to extract a basis for the subspace of bright states: we simply select those eigenstates with at least one component in the measured region $A$ as defined in Eq. \eqref{eq:bright_subspace}. These are: i) $\ket{N}$ because it is spread over the whole lattice; ii) $\ket{0,l}$ with $l=m,\dots,N-1$ as seen from Eq. \eqref{eq:normalized_degenerate_states}. Explicitly, the bright subspace is given by
\begin{equation}\label{eq:bright_states_set}
    B = \Span\left\{\ket{N}\,\,,\,\,\quad \ket{0, l}\quad l=m,\dots,N-1\right\} \quad .
\end{equation}
As we already mentioned above, we now calculate the action of the effective time evolution operator $\tilde U_\tau$ on each of these basis states one by one and at the sum over all of them. We start with the easiest case, namely, the non-degenerate state $\ket{N}$, followed by the degenerate states $\ket{0,l}$ for $l=m,\dots, N-1$.

\vskip 0.3cm

\noindent{\bf Non-degenerate eigenspace:}
We start setting $\ket{\psi_0}=\ket{\Lambda =N}$ given in Eq. \eqref{eq:lambda_non_degenerate}.  
We recall that the measurement protocol is that described in Section \ref{section:main_concepts}. As in that section, we use again the short hand notation $\ket{\psi^\pm_{\tau_i}}\equiv \ket{\psi^\pm_i}$ defined in Eq. \eqref{eq:psi_minus} and Eq. \eqref{eq:psi_plus}. We also abbreviate $U_{\tau_i}\equiv U_i$ where $U_\tau$ is defined in Eq. \eqref{eq:unitary_evolution_mean_field}. During the first time step we get
\begin{equation}
    \ket{\psi_1^+} = P_{A^{\perp}}U_1 \ket{N} = e^{i\tau_1 JN}P_{A^{\perp}}\ket{N}
\end{equation}
because $\ket{N}$ is eigenstate of $H$ with eigenvalue $N$ (see Eq. \eqref{eq:H_ata}).
Then we have
\begin{align}
    \ket{\psi_2^+} = P_{A^{\perp}}U_2 \ket{\psi_1^+} &= e^{i\tau_1 JN}P_{A^{\perp}}U_2P_{A^{\perp}}\ket{N}
    \nonumber
    \\
    & = e^{i\tau_1 J N}(1+m b_2)P_{A^{\perp}}\ket{N} \label{eq:second_time_step}
\end{align}
where $b_i\equiv b_{\tau_i}$ defined in Eq. \eqref{eq:definition_b_mean_field}. The second line follows from the following two facts: first, recalling tha the matrix $E$ is the matrix with all $1$s and $P_{A^\perp}$ is the projector onto the sites from $1$ to $m$ we have that
\begin{equation}
    P_{A^{\perp}}E P_{A^{\perp}} = \begin{pmatrix}
        E_{m\times m} & 0\\
        0 & 0
    \end{pmatrix}
\end{equation}
where $E_{m\times m}$ is the matrix with all $1$s of dimension $m\times m$. This implies that 
\begin{equation}\label{eq:projector_on_giant_eigenstate}
    P_{A^{\perp}}E P_{A^{\perp}}\ket{N} = m P_{A^{\perp}}\ket{N}\quad .
\end{equation}
Second, using the decomposition $U_i=1+b_i E$  (see Eq. \eqref{eq:definition_b_mean_field}) gives Eq. \eqref{eq:second_time_step}. Then repeating this argument $n$ times we arrive at
\begin{align}\label{eq:ndeg_sub_psi0}
    \ket{\psi_n^+} =  e^{i\tau_1 JN}\prod_{j=2}^n(1+mb_j)P_{A^{\perp}}\ket{N}\,.
\end{align}
Plugging this result in Eq. \eqref{eq:survival_conditioned} and recalling that $\ket{\psi_0}=\ket{N}$ gives
\begin{align}\label{eq:surv_sqrtN}
    \|\tilde U_n\dots \tilde U_1\ket{\psi_0}\|^2 &= \braket{\psi_n^+|\psi_n^+} =\prod_{j=2}^n|1+mb_j|^2\braket{N |P_{A^{\perp}}|N}
    \nonumber
    \\
    & =\prod_{j=2}^n|1+mb_j|^2\frac{m}{N}\,.
\end{align}
Here, in the last line we have used $\braket{N|P_{A^{\perp}}|N}= \frac{m}{N}$ (which in turn follows from of Eq. \eqref{eq:lambda_non_degenerate}) and the fact that
\begin{equation}
    P_{A^{\perp}}\ket{N} = \frac{1}{\sqrt{N}}\begin{pmatrix}
        &\vec{1}_m  \\ &\vec{0}_{N-m}
    \end{pmatrix}
\end{equation}
where $\vec{1}_m$ is the column vector with all $1$ of dimension $m$ and $\vec{0}_{N-m}$ the column vector of dimension $N-m$ of all $0$.
We note that, in this simple case where the initial state is just $\ket{\psi_0}=\ket{N}$, the Laplace transform of the non-deteciton probability $\hat S(s)$ in Eq. \eqref{eq:surv_lapla_transf} takes a particularly simple form. Indeed, inserting Eq. \eqref{eq:surv_sqrtN} into the expression for the Laplace transform of the non-detection probability in Eq. \eqref{eq:surv_lapla_transf} and summing the resulting geometric series we obtain
\begin{equation}\label{eq:prediction_lambda_bright}
    \hat S(s) = \frac{1}{r+s}\left[1 + \frac{m}{N}\frac{\frac{r}{r+s}}{1-\overline{|\Lambda|^2}}\right]
\end{equation}
where we have used
\begin{equation}
    \frac{1-\hat f (s)}{s}=\frac{1}{r+s}
\end{equation}
valid for the Poissonian protocol where $f(\tau) = r e^{-r \tau}$  and defined
\begin{equation}\label{eq:lambda_defintion}
    \Lambda \equiv \Lambda(\tau) = 1+mb_\tau
\end{equation}
with $b_\tau$ from Eq. \eqref{eq:definition_b_mean_field}. Also, for a function $g(\tau)$ we use the definition \begin{equation}\label{eq:overbar_definition}
    \overline{g}(s) \equiv \int_0^\infty \dd \tau\, e^{-s \tau } g(\tau)f(\tau)\quad \text{where} \quad f(\tau) = r \,e^{-r\tau}\,.
\end{equation} 
In particular, for the Poissonian protocol and it is easy to calculate
\begin{equation}\label{eq:lambda_laplace_transform}
    \overline{|\Lambda|^2} =\frac{r \left(J^2 \left(2 m^2-2 m N+N^2\right)+(r+s)^2\right)}{(r+s) \left(J^2 N^2+(r+s)^2\right)}
\end{equation}
which implies
\begin{equation}\label{eq:common_factor_lambda_squared}
    \frac{1}{1-\overline{|\Lambda|^2}} = \frac{(r+s) \left(J^2 N^2+(r+s)^2\right)}{J^2 \left(-2
   m^2 r+2 m N r+N^2 s\right)+s (r+s)^2}\quad .
\end{equation}

\vskip 0.3cm

\noindent{\bf Degenerate eigenspace:}
Now we calculate the repeated action of the effective time evolution operator $\tilde U_t$ on the degenerate bright eigenstates. According to Eq. \eqref{eq:bright_states_set} these are $\ket{0,l}$ for $l=m,\dots,N-1$ whose explicit form is in Eq. \eqref{eq:normalized_degenerate_states}. 
As before, after the first time step and the first measurement we have
\begin{equation}
    \ket{\psi_1^+} = P_{A^{\perp}}U_1 \ket{0,l} = P_{A^{\perp}}\ket{0,l}
\end{equation}
because these are eigenstates with zero eigenvalue of the Hamiltonian. After the second step
\begin{align}
    \ket{\psi_2^+} = P_{A^{\perp}}U_2 \ket{\psi_1^+} &= P_{A^{\perp}}U_2P_{A^{\perp}}\ket{0,l}\quad .
\end{align}
Thus we compute
\begin{align}\label{eq:projected_normalized_eigenstates}
    P_{A^{\perp}}\ket{0,l} = P_{A^{\perp}}C_l \begin{pmatrix}
        -1\\ \vdots\\-1 \\ l \\ 0 \\ \vdots\\0 
    \end{pmatrix}
    =C_l\begin{pmatrix}
        -1\\ \vdots\\-1 \\ 0 \\ 0 \\ \vdots\\0 
    \end{pmatrix} = -\sqrt{N}C_l P_{A^{\perp}}\ket{N}
\end{align}
where $C_l$ as in Eq. \eqref{eq:normalized_degenerate_states} is the normalization factor of the states and we used the fact that the first $l$ components are $1$ and only the $l+1$ is $l$ and $l\geq m$ while $A^{\perp}=[1,m]$; the last equality follows from the definition of $\ket{N}$.
In this way
\begin{align}\label{eq:projector_on_eigenstates}
    \ket{\psi^+_2}=P_{A^{\perp}}U_2\ket{\psi_1^+}=P_{A^{\perp}}U_2P_{A^{\perp}}\ket{0,l}&=P_{A^{\perp}}(1+b_2 E)P_{A^{\perp}}\ket{0,l}
    \nonumber  
    \\
    &= -\sqrt{N }C_lP_{A^{\perp}}\ket{N}-\sqrt{N}C_l P_{A^{\perp}}b_2 E P_{A^{\perp}}\ket{N}
    \nonumber  
    \\ 
    &=-\sqrt{N }C_lP_{A^{\perp}}\ket{N}-\sqrt{N}C_l m b_2 P_{A^{\perp}}\ket{N}
    \nonumber 
    \\  
    &=-\sqrt{N }C_l\left(1+ m b_2\right) P_{A^{\perp}}\ket{N}\quad .
\end{align}
To arrive at the last line we proceeded in the following way: in the first line we used $U_i=1+b_iE$ in Eq. \eqref{eq:definition_b_mean_field}; in going from the first to the second line we used the result Eq. \eqref{eq:projected_normalized_eigenstates}; from the second to the third we used Eq. \eqref{eq:projector_on_giant_eigenstate}; in the fourth we recombined the terms.
Repeating for $n$ steps we obtain
\begin{equation}\label{eq:deg_sub_psi0}
   \ket{\psi_n^+} =  (-\sqrt{N}C_l)\prod_{j=2}^n(1+mb_j)P_{A^{\perp}}\ket{N}\,.
\end{equation}
In this case too, Eq. \eqref{eq:surv_lapla_transf} takes a simple form. Using $\braket{N|P_{A^{\perp}}|N}=m/N$ we obtain, after averaging and summing the geometric series in Eq. \eqref{eq:surv_lapla_transf}
\begin{equation}
    \hat S(s) =\frac{1}{r+s}\left[1 + \frac{m}{N}\frac{N\frac{r}{r+s}|C_l|^2}{1- \overline{|\Lambda}|^2}\right]
\end{equation}
where $\overline{|\Lambda|^2}$ is defined in Eq. \eqref{eq:lambda_laplace_transform} and $C_l$ in Eq. \eqref{eq:normalized_degenerate_states}.

\vskip 0.3cm

\noindent{\bf General solution in the bright subspace:}
It is now simple to consider a general bright state as in Eq. \eqref{eq:bright_state_expansion}. Using Eq. \eqref{eq:ndeg_sub_psi0} and Eq. \eqref{eq:deg_sub_psi0} we find
\begin{align}
 \ket{\psi_n^+} =  e^{i\tau_1 JN}\prod_{j=2}^n(1+mb_j)P_{A^{\perp}}\ket{N}\braket{N|\psi_0}    + \sum_{l=m}^{N-1}(-\sqrt{N}C_l)\prod_{j=2}^n(1+mb_j)P_{A^{\perp}}\ket{N}\braket{0,l|\psi_0}\,.
\end{align}
from which, extracting the norm squared and inserting in Eq. \eqref{eq:surv_lapla_transf} results in
\begin{align}\label{eq:laplace_transform_survival}
    \hat S(s) = \frac{1}{r+s}\Bigg\{1  &+ \frac{m}{N}\frac{1}{1-\overline{|\Lambda|^2}}\Bigg[\frac{r}{r+s}|\braket{N|\psi_0}|^2 + N\frac{r}{r+s}\left|\sum_{l=m}^{N-1}C_l\braket{0,l|\psi_0}\right|^2 
    \nonumber
    \\
    &- 2 \sum_{l=m}^{N-1}\Re \left(\overline{e^{-i\tau J N}}\sqrt{N}C_l\braket{0,l|\psi_0}\braket{\psi_0|N}\right)\Bigg]\Bigg\}\,.
\end{align}
This is the general expression for an initial state \emph{lying} completely in the bright subspace. We can seefrom the three pieces all the eigenstates are equally equally important and contribute to the $\hat S(s)$.

\section{Optimal detection}\label{sec:optimal_detection}
Now that we have the general expression for the Laplace transform $\hat S(s)$ of the non-detection probability Eq. \eqref{eq:laplace_transform_survival}, we can investigate what are the conditions on the initial bright state $\ket{\psi_0}$ for optimal detection by computing the MFDT and the first detection probability. We will compute exactly the MFDT $T(r)$ and show that it has a minimum at a finite $r^*$ depending on the spacial localization of the initial state. Then, we will calculate the first detection probability $F(t)$ and analyze its asymptotic behaviors for large and short times.  

As a remark on notation we define the two coefficients
\begin{equation}\label{eq:cAdef}
    c_A =\sum_{x\in A}\braket{x|\psi_0}= \sum_{x\in A}\psi_0(x)\quad , \quad c_{A^\perp} = \sum_{x\in A^\perp}\braket{x|\psi_0}=\sum_{x\in A^\perp}\psi_0(x)\,.
\end{equation}
These are the sums of the coordinates of the initial state restricted to the target set $A$ and its complement $A^\perp$. These two quantities play an important role in the upcoming calculations.
\subsection{Mean first detection time}\label{subsec:mfdt}
We now calculate the full expression of the mean first detection time as a function of the measurement rate $r$ for any initial bright state. In order to simplify Eq. \eqref{eq:laplace_transform_survival} we will use two non-trivial sum rules that we have calculated in Appendix \ref{app:scalar_products}: these are Eq. \eqref{eq:tot_sum_rule1} and Eq. \eqref{eq:tot_sum_rule2}. First we look at the interference term on the left of Eq. \eqref{eq:laplace_transform_survival}. We write
\begin{equation}
    \overline{e^{-\ii J \tau N}} = \overline{\cos(J N \tau)} + \ii\, \overline{\sin(J N \tau)}\,.
\end{equation}
Using that for $z\in \C$ and $x,y\in \R$ we have $\Re((x + \ii y)z)= x \Re z - y \Im z$ and the two sum rules Eq. \eqref{eq:tot_sum_rule1} and Eq. \eqref{eq:tot_sum_rule2} to perform the sum over $l$ we can compute 
\begin{align}
    &\sum_{l=m}^{N-1}\Re \Bigg\{\overline{e^{-i\tau J N}}\sqrt{N}C_l\braket{0,l|\psi_0}\braket{\psi_0|N}\Bigg\}  \nonumber
    \\
    &= \frac{1}{N}\left(|c_A|^2-|c_{A^{\perp}}|^2\frac{N-m}{m} \right)\overline{\cos(\tau J N)} +\frac{1}{N}\Re\left\{ \overline{e^{-iJ N \tau}} \left[-c_{A^{\perp}}c_{A}^*\frac{N-m}{m}+ c^*_{A^{\perp}}c_A\right]\right\}
    \nonumber
    \\
    & =\frac{1}{N}\left(|c_A|^2-|c_{A^{\perp}}|^2\frac{N-m}{m} \right)\overline{\cos(NJ\tau)}
     +\frac{1}{N} \Re\left\{  -c_{A^{\perp}}c_{A}^*\frac{N-m}{m}+ c^*_{A^{\perp}}c_A\right\} \overline{\cos(NJ\tau)}
     \nonumber
     \\
     &
     -\frac{1}{N}  \Im\left\{-c_{A^{\perp}}c_{A}^*\frac{N-m}{m}+ c^*_{A^{\perp}}c_A\right\}\overline{\sin(NJ\tau)}\label{eq:interference_generic_case}\,.
\end{align}
The remaining terms in Eq. \eqref{eq:laplace_transform_survival} are
\begin{align}\label{eq:non_degenerate_generic_case}
    |\braket{N|\psi_0}|^2  = \frac{1}{N}\left[|c_A|^2 + |c_{A^\perp}|^2 + 2 \Re \left\{c_A c^*_{A^\perp}\right\}\right] 
\end{align}
computed thanks to Eq. \eqref{eq:tot_sum_rule1} and analogously,
\begin{align}\label{eq:degenerate_states_generic_case}
    \left|\sum_{l=m}^{N-1}C_l\braket{0,l|\psi_0}\right|^2 = \frac{1}{N^2}\left[|c_A|^2 + \left(\frac{N-m}{m}\right)^2 |c_{A^\perp}|^2 - 2 \frac{N-m}{m}\Re\left\{c_A c^*_{A^\perp}\right\}\right]\quad.
\end{align}
computed using Eq. \eqref{eq:tot_sum_rule2}.
Plugging Eq. \eqref{eq:interference_generic_case}, Eq. \eqref{eq:non_degenerate_generic_case} and Eq. \eqref{eq:degenerate_states_generic_case} into the expression for  $\hat S(s)$ in Eq. \eqref{eq:laplace_transform_survival}, we obtain
\begin{equation}\label{eq:non_det_general}
    \hat S(s) =  \frac{1}{r+s} \Bigg\{1  + \frac{1}{1-\overline{|\Lambda|^2}}\Bigg[a_1\frac{r}{r+s} + a_2\overline{\cos(NJ\tau)}  
   + a_3\overline{\sin(NJ\tau)}
   \Bigg]\Bigg\}\,.
\end{equation}
The coefficients in Eq. \eqref{eq:non_det_general} are functionals of the initial state and read as follows
\begin{subequations}\label{eq:coefficients}
    \begin{equation}
        a_1 =  \frac{m}{N^2}\left\{2|c_A|^2 + |c_{A^\perp}|^2\left[1+\left(\frac{N-m}{m}\right)^2\right] + 2 \Re \left\{c_A c^*_{A^\perp}\right\}\left[1-\frac{N-m}{m}\right] \right\} \quad ,
    \end{equation}
    \begin{equation}
        a_2 =-2\frac{m}{N^2}\left(|c_A|^2-|c_{A^{\perp}}|^2\frac{N-m}{m} \right) -2 \frac{m}{N^2}\Re \left\{c_A c^*_{A^\perp}\right\}\left[1-\frac{N-m}{m}\right] \quad ,
    \end{equation}
    \begin{equation}
        a_3 =  2\frac{m}{N^2}\Im\left\{ c^*_{A^{\perp}}c_A\right\}\left[1+\frac{N-m}{m}\right]
    \end{equation}
\end{subequations}
which can be obtained after some simple algebra involving complex numbers. These coefficients obey the simple sum rule
\begin{equation}\label{eq:sum_a1_a2}
    a_1 + a_2= |c_{A^\perp}|^2/m\,.
\end{equation}
Furthermore, recalling the definition in Eq. \eqref{eq:overbar_definition}, we can compute
\begin{equation}\label{eq:laplace_sine_cosine}
    \overline{\cos(JN\tau)} = \frac{r(r+ s)}{(r+ s)^2+(JN)^2}\quad , \quad \overline{\sin(JN\tau)} = \frac{ rJN}{(r+ s)^2 + (JN)^2}\quad .
\end{equation}
Substituting Eq. \eqref{eq:common_factor_lambda_squared} and Eq. \eqref{eq:laplace_sine_cosine} in Eq. \eqref{eq:non_det_general}, after some algebra we obtain the exact Laplace transform of the non-detection probability
\begin{align}\label{eq:non_det_poiss}
    \hat S(s) = \frac{1}{r+s} \Bigg\{1  + \frac{(r+s) \left(J^2 N^2+(r+s)^2\right)}{J^2 \left(-2
   m^2 r+2 m N r+N^2 s\right)+s (r+s)^2}&\Bigg[a_1\frac{r}{r+s} + a_2\frac{r(r+ s)}{(r+ s)^2+(JN)^2} 
   \nonumber
   \\
   &+ a_3\frac{rJN}{(r+ s)^2+(JN)^2}
   \Bigg]\Bigg\}\,.
\end{align}
This expression demonstrates that all the dependence of the initial state as well as the detailed quantum mechanical interfence pattern of different parts of the state are encoded in the coefficients $a_i$.
Setting $s=0$ in Eq. \eqref{eq:non_det_poiss} gives the exact MFDT as a function of the measurement rate
\begin{align}\label{eq:mfdt_exact}
    T(r) = \frac{1}{r} \Bigg\{1  &+ \frac{r \left(J^2 N^2+r^2\right)}{J^2 \left(-2
   m^2 r+2 m N r\right)}\Bigg[a_1 + a_2\frac{r^2}{r^2+ (J N)^2}  
   + a_3\frac{ rJN}{r^2 + (JN)^2} 
   \Bigg]\Bigg\}\,.
\end{align}
This exact result is compared to numerical simulations in the left panel of Fig. \ref{fig:mfdt}.
To study the asymptotic behavior as a function of $r$ we note that if $c_{A^\perp}=0$, i.e., the initial state is is completely localized within the target subspace $A$, then from Eq. \eqref{eq:sum_a1_a2} $a_1=-a_2$. In all the other cases $a_1 \neq a_2$. 
With these observations, it is a simple matter to derive the asymptotic behavior summarized in Eq. \eqref{eq:MFDT_asymptotic_large} and Eq. \eqref{eq:MFDT_asymptotic_small}. 
Differentiating Eq. \eqref{eq:mfdt_exact} with respect to $r$ one finds that there is an optimal rate $r^*$ that minimizes $T(r)$ given explicitly by
\begin{equation}\label{eq:rstar}
    r^*= \sqrt{m}\frac{\sqrt{J^2 N^2 a_1 + 2 J^2 N m - 2 J^2 m^2}}{|c_{A^\perp}|}\,.
\end{equation}
From this formula, it is clear that $r^*\to+\infty$ when $c_{A^\perp}\to 0$, i.e., the initial state is entirely localized in the measured region $A$.

\subsection{First detection probability}
To compute the first detection probability $F(t)$ as a function of time for a given rate $r$ we use the relation between the Laplace transforms of these quantities given in Eq. \eqref{eq:laplace_first_det}. 
Plugging Eq. \eqref{eq:non_det_poiss} in Eq. \eqref{eq:laplace_first_det} and after simple algebra we obtain the generating function of the first detection time
\begin{equation}\label{eq:F_hat_lapl}
    \hat F(s) =1- \frac{s}{r+s} -\frac{r s \left[a_1 \left(J^2 N^2+(r+s)^2\right)+(r+s)
   \left(a_3 J N+a_2 (r+s)\right)\right]}{(r+s) \left[J^2
   \left(-2 m^2 r+2 m N r+N^2 s\right)+s
   (r+s)^2\right]}\,.
\end{equation}
It is easily checked that this is correctly normalized $\hat F(0)=1$. For complex $s$, this function is meromorphic, i.e., the only singularities are poles.
It can be inverted using the inversion formula for the Laplace transform in combination with residues theorem as
\begin{equation}\label{eq:inversion_F}
    F(t) = \int_{\gamma-\ii \infty}^{\gamma + \ii \infty}\frac{\dd s}{2\pi \ii} e^{st}\hat F(s) = \sum_i \Res \left(e^{s t}\hat F(s), s=s_i\right) \quad , \quad t>0
\end{equation}
where $\gamma$ is chosen such that all the poles of $\hat F(s)$ lie on the left half plane $\Re s\leq \gamma$ and $s_i$ are the poles of the denominator.
An important observation is that the poles of the denominator are independent from the coefficients $a_i$. Since the $a_i$ encode all the information about the initial state (see Eq. \eqref{eq:coefficients}), the large time behavior of $F(t)$ will not depend on it. Physically this means that the way the system dephases and decays is a property of the system and not of the particular state. 
To analyze the poles, it is convenient to define the polynomials 
\begin{equation}
    P(s) = s r\left[a_1  \left(J^2 N^2+(r+s)^2\right)+ (r+s) \left(a_3 J N+a_2 (r+s)\right)\right]
\end{equation}
\begin{equation}
    D(s) =(s+r)Q(s)
\end{equation}
associated to the numerator and the denominator of the rational part of Eq. \eqref{eq:F_hat_lapl}. Here we use the additional definition 
\begin{align}
    Q(s) &= 2J^2m\,r(N-m)+s \left(J^2 N^2+r^2\right)+2 r
   s^2+s^3
   \nonumber
   \\
   & = (s-s_1)(s-s_2)(s-s_3)
\end{align}
 We notice that both polynomials, $P$ and $D$, have real coefficients implying that for any complex $s$ we have  $P(s^*)=P^*(s)$ and $D(s^*)=D^*(s)$. One zero of $D(s)$ is simply $s_4 = -r$. We discuss below that $Q(s)$ has  three additional zeros, one real that we call $s_1$ and a complex one $s_2 = s_R + \ii s_I$ with its conjugate $s_3=s_2^*$.

In this way, we can formally write the full solution
\begin{align}\label{eq:exact_F}
    F(t) = r e^{-r t}+\frac{P(-r)}{(r+s_1)|r+s_2|^2}e^{-rt} &- \frac{P(s_1)}{(s_1+r)|s_1-s_2|^2}e^{s_1 t} 
    \nonumber
    \\
    &- 2e^{s_R t}\Re\left\{e^{\ii s_I t}\frac{P(s_2)}{2\ii s_I(s_2-s_1)(s_2+r)}\right\} 
\end{align}
where we have used that the denominator factors as $D(s) = (s-r)(s-s_1)(s-s_2)(s-s_3)$ and the fact that all the $s_i$ are simple zeros to compute residues according to the standard formula. Below we compute explicitly $s_1,s_2,s_3$ in terms of the system parameters. This analytical result is compared to exact numerical simulations in Fig. \ref{fig:F_t}.
We now discuss the nature of the zeros. The denominator vanishes at $s_4=-r$ and when
\begin{equation}\label{eq:poly_den}
   Q(s) =2J^2m\,r(N-m)+s \left(J^2 N^2+r^2\right)+2 r
   s^2+s^3=0\,.
\end{equation}
Clearly, $Q(s)$ has three roots: one real root $s_1$ and two complex conjugate roots $s_2=s_3^* = s_R + \ii s_I$, with $s_I\neq 0$. In Appendix \ref{app:hurwitz} we show that the real root $s_1$ satisfies the bounds (see Eq. \eqref{eq:rlesss1} and Eq. \eqref{eq:sRlesss1})
\begin{equation}\label{eq:ineq_roots}
    -r<s_1<0\quad , \quad s_R<s_1<0\,.
\end{equation}
The polynomial $Q(s)$ is a cubic and it can be transformed into a depressed cubic, i.e., a cubit without the quadratic term, by the standard transformation 
\begin{equation}
    s=z-2r/3\label{eq:z_shift}
\end{equation}
obtaining
\begin{equation}\label{eq:depressed_poly}
    -2 J^2 m^2 r+2 J^2 m N r-\frac{2}{3} J^2 N^2 r-\frac{2
   r^3}{27}+z \left(J^2
   N^2-\frac{r^2}{3}\right)+z^3=0\,.
\end{equation}
For a polynomial of the type
\begin{equation}\label{eq:depressed_poly_ref}
    z^3 + p z + q = 0
\end{equation}
the discriminant is
\begin{equation}
    \Delta = \left(\frac{p}{3}\right)^3 + \left(\frac{p}{2}\right)^2\,.
\end{equation}
Comparing the coefficients in Eq. \eqref{eq:depressed_poly} to those in Eq. \eqref{eq:depressed_poly_ref} we can identify
\begin{equation}\label{eq:q_coeff}
    q=-2 J^2 m^2 r+2 J^2 m N r-\frac{2}{3} J^2 N^2 r-\frac{2
   r^3}{27}
\end{equation} 
\begin{equation}\label{eq:p_coeff}
    p = J^2
   N^2-\frac{r^2}{3}\,.
\end{equation}
There is one real root and one complex conjugate pair if and only if $\Delta>0$. In Appendix \ref{app:hurwitz} we show that this is the case for all $N\geq 1$ and $r>0$ 
irrespectively from $J$. Let us call $z_1$ the real root and $z_2=z^*_3$ the conjugate roots of the depressed cubic Eq. \eqref{eq:depressed_poly}.
Cardano's fomulas tell that the real root is simply
\begin{equation}\label{eq:real_root}
    z_1 = u + v
\end{equation}
where 
\begin{equation}\label{eq:cardano}
    u = \sqrt[3]{- \frac{q}{2} + \sqrt{\Delta}}\quad , \quad v = \sqrt[3]{- \frac{q}{2} - \sqrt{\Delta}}
\end{equation}
while the complex root is given by
\begin{equation}\label{eq:conjugate_roots}
    z_2 = -\frac{1}{2}(u+v) + \frac{\ii\sqrt{3}}{2}(u-v)\equiv z_R + \ii z_I\,.
\end{equation}
Using Eq. \eqref{eq:z_shift} and Eq. \eqref{eq:real_root} we can the real root of $Q(s)$ in Eq. \eqref{eq:poly_den} as
\begin{equation}\label{eq:roots1}
    s_1 = u+v -\frac{2}{3}r
\end{equation}
with $u,v$ given by Eq. \eqref{eq:cardano}. 
This analysis allows us to conclude that for $t\to + \infty$
\begin{equation}
    F(t) \sim \frac{P(s_1)}{|s_1-s_2|^2}e^{ -\frac{t}{t_m(r)}} \quad ,\quad \text{for } r\to 0  \quad  \quad \text{with } t_m(r) = \frac{1}{|s_1|}
\end{equation}
because $t_m$ is the largest time scale due to Eq. \eqref{eq:ineq_roots}. Here, $s_1$ in given explicitly in Eq. \eqref{eq:roots1} and it is this expression that is used in the right plot of Fig. \ref{fig:mfdt}. The time scale $t_m(r)$ has the physical meaning of the 'maximal time` one has to wait before we detect the target $A$ \cite{Kulkarni_2023}.

To study the opposite $t\to 0$ limit one simply looks at the large $s$ behavior of the generating function $\hat F(s)$. Indeed, if $\hat F(s)\sim s^{-\mu-1}$ for $s\to +\infty$ then $F(t)\sim t^\mu$ for $t\to 0$.

Expanding Eq. \eqref{eq:F_hat_lapl} up to third order one finds
\begin{align}\label{eq:F_hat_lapl_small_s}
    \hat F(s) = \frac{r(1-a_1-a_2)}{s}&+\frac{-a_3 J N r+r^2(a_1+a_2-1)}{s^2}
    \nonumber
    \\
    &+\frac{a_2 J^2 N^2 r+2 a_3 J N r^2+r^3(1-a_1-a_2)}{s^3}+O\left(\frac{1}{s^4}\right) \,.
\end{align}
From this expression it looks like that the behavior strongly depends on the coefficients $a_i$ and so on the initial state. Using Eq. \eqref{eq:sum_a1_a2},  the vanishing of the leading term  $\propto s^{-1}$ in Eq. \eqref{eq:F_hat_lapl_small_s} is equivalent to non-trivial sum rule
\begin{equation}\label{eq:constraint_cAperp}
    |c_{A^\perp}|^2 = m
\end{equation}
where $m$ is the size of $A^\perp$.
Recalling the definition of $c_{A^\perp}$ as the sum of the coordinates belonging to $A^\perp$ of the initial state in Eq. \eqref{eq:cAdef}, this fact implies that the constraint Eq. \eqref{eq:constraint_cAperp} can only be satisfied the initial state is given by the special state $\ket{\psi^*} = m^{-1/2}P_{A^\perp}\ket{N}$ (see Eq. \eqref{eq:special_state}), where $\ket{N}$ is the energy eigesntate spread all over the lattice in Eq. \eqref{eq:lambda_non_degenerate}.
 When this happens then $|c_A|=\left|\sum_{x\in A}\psi_0(x)\right|=0$ because the state has no components in $A$ and from Eq. \eqref{eq:coefficients} we see that $a_3=0$ and $a_2 = 2 \frac{N-m}{N^2}$. 
Using these observations in Eq. \eqref{eq:F_hat_lapl_small_s}, we conclude that for this special initial state $\ket{\psi^*}$, we have that
\begin{equation}
    \hat F(s) \overset{s\to \infty}{\sim} 2J^2(N-m)r\,s^{-3} \implies F(t) \overset{t\to 0}{\sim} J^2 (N-m)r t^2
\end{equation} 
In all the other cases we instead obtain $F(t)\sim \text{const}.$ as reported in Eq. \eqref{eq:asy_F_short}. 
This is the generalization of the behavior observed in the single q-bit case \cite{Kulkarni_2023} for which $N-m=1$.

\section{Conclusions}\label{section:conclusions}

To summarize, in this work we studied the quantum mechanical problem of optimal detection of a quantum state. The system evolves under the unitary time evolution ruled by the Schr\"odinger equation and it is probed at random times $T_i$ with rate $r$ for a time of duration $t$. 

Our goal was to understand under what conditions the times scales associated to the detection protocol could be optimized, i.e., made minimal. We have concentrated on two time scales: the mean first detection time $T(r)$ and the time scale governing the exponential decay of the first detection probability $t_m(r)$.
As a simple model we considered a quantum particle on a one-dimensional lattice with an all-to-all Hamiltonian and as a target detection subspace a set of continuous $N-m$ sites. This setup allowed for exact analytical solutions for both the mean first detection time $T(r)$ and the first detection probability $F(t)$ and we find that both depend non-trivially on the  measurement rate $r$, the target subspace $A$ and the initial state $\ket{\psi_0}$.

We confirm the significant role of dark states \cite{krovi_2006, krovi_2006b, Thiel_2020} in quantum search processes. These are unique to quantum mechanics and, when present, prevent perfect detection. As a consequence,  our analysis focuses on initial states that are bright, meaning they have no component in the dark subspace, ensuring that the mean first detection time is finite.

For this fully-connected Hamiltonian, we derived an explicit formula for $T(r)$. Importantly, we found that a \emph{finite optimal resetting rate} $r^*$ that minimizes $T(r)$ exists only when the initial state has some overlap with the unmeasured subspace $A^\perp$. If the initial state is entirely localized within the target subspace $A$, the optimal strategy involves measuring at an infinite rate, leveraging the quantum Zeno effect \cite{misra1977, itano1990}. This divergence at $r \to +\infty$ when the initial state overlaps with $A^\perp$ (and thus a finite $r^*$ exists) highlights a fundamental difference in optimal detection strategies based on the initial quantum state.

Beyond the mean time, we also thoroughly characterized the first detection probability $F(t)$, i.e., the probability that the first detection occurs before time $t$. This function exhibits an initial power-law behavior, which we show is \emph{highly dependent on the initial state}. Specifically, for a general initial state, $F(t) \sim \text{const.}$ as $t \to 0$. However, for a special initial state completely localized in $A^\perp$, we find $F(t) \sim r(N-m)t^2$, indicating a slower onset of detection probability.

In contrast, at large times, $F(t)$ decays exponentially as $F(t)= \exp(-t/t_m(r))$ for any initial state. The characteristic timescale $t_m(r)$ also exhibits a minimum at an optimal rate $r^*_m$. Crucially, this optimal rate $r^*_m$ exists for all \emph{bright} initial states, irrespective of their overlap with the unmeasured subspace. This independence from the initial state at long times can be attributed to the decoherence induced by the measurements, effectively erasing the memory of the initial state.

Our results provide a rare instance of an analytically solvable model for monitored quantum dynamics, offering valuable insights into the fundamental mechanisms governing optimal detection in quantum systems. The dependence of optimal detection strategies on the initial state and the existence of dark states are unique features of quantum mechanics, setting it apart from classical random search processes. We believe that this work contributes to a deeper understanding of how to improve quantum search and has potential in the field of quantum technologies. Further research could explore the implications of these findings for more complex quantum systems and different measurement protocols. For example, it would be interesting to analyze more general models, like particles hopping on different graphs, higher dimensional systems and many-body effects. Another interesting direction would be to compare different measurement protocols and see how their performances differ from one another, something which we have decided not to do in the present work to avoid masking our main results. Quantum generalization of standard \cite{VVS07} and new \cite{delvecchio2025b} search protocols based on different types of information are also an interesting future direction.

\begin{appendices}
\section{The role of degeneracy}\label{app:degeneracy}

In this short Appendix we discuss the presence of degeneracy in the spectrum of the Hamiltonian $H$ in relation to dark states.
When $H$ has non-trivial degenerate eigenspaces the characterization of dark states becomes more subtle \cite{krovi_2006, verbanov_2008, Thiel_2020}. Indeed, when the spectrum is non-degenerate, to check if each of the eigenstates satisfies $P_A \ket{\mu}=0$, according to Eq. \eqref{eq:dark_subspace2}. This would suffice to say that they are dark.

Suppose now there is an entire eigenspace with energy $\mu$ and degeneracy $g>1$. Let us consider a basis for this eigenspace $\{\ket{\mu_i}\}_{i=1}^g$ where $H \ket{\mu_i}=\mu \ket{\mu_i}$ for $i=1,\dots,g$. Since the Hamiltonian $H$ is hermitian, different eigenspaces are orthogonal to each other but eigenstates within the same eigenspace are not necessarily orthogonal between them \cite{artin2011algebra}. 

Nevertheless, it is possible to find infinitely many orthonormal bases of eigenstates for this degenerate eigenspace via some kind of orthogonalization procedure such as the Gram-Schmidt method or the $QR$ decomposition. Any linear combination of the eigenstates will still be an eigenstate with the same eigenvalue. This simple fact means that, if there is sufficient degeneracy, it might be possible to construct linear combinations of eigenstates within that eigenspace that have zero overlap with the target subspace $A$ we are trying to measure. This linear combination will then be a dark eigenstate because in taking linear combinations of eigenstates with the same eigenvalue we never leave the degenerate eigenspace. 
Within all the orthonormal eigenbases of eigenstates associated to this degenerate eigenspace there is only one which reveals dark states and it can be found algorithmically as follows. 

We take $\ket{\psi} = \sum_{m=1}^g a_m \ket{\mu_m}$ as a linear combination of $g> 1$ degenerate eigenstates $\{\ket{\mu_i}\}_{i=1}^g$ all with eigenvalue $\mu$. To construct a dark state we need to find the coefficients $a_i$ such that this state has no component in $A$, i.e., solve the linear system
\begin{equation}\label{eq:deg_dark}
    P_A \ket{\psi} = \sum_{m=1}^g a_m P_A\ket{\mu_m}=0\quad, \quad \sum_{m=1}^g |a_m|^2 = 1
\end{equation}
where $\ket{\mu_m}$ are the $g$ degenerate eigenstates and $P_A$ the projector into the target set $A$. In matrix form we need to solve
\begin{equation}\label{eq:deg_dark_matrix}
    T \vec{a}= 0
\end{equation}
where
\begin{equation}\label{eq:T_matrix}
    T = \begin{pmatrix}
        \vline & \vline &  & \vline \\
        P_A \ket{\mu_1} & P_A \ket{\mu_2} & \cdots & P_A \ket{\mu_g}\\
        \vline & \vline &  & \vline 
        \end{pmatrix}\quad \text{and}\quad \vec{a}=\begin{pmatrix}
        a_1 \\ \vdots \\a_g 
    \end{pmatrix}
\end{equation}
The coefficients $\{a_i\}_{i=1}^g$ in Eq. \eqref{eq:deg_dark} that make the linear combination vanish belong to the kernel of the $N' \times g$ matrix $T$ having columns $P_A \ket{\mu_i}$ for $i=1,\dots ,g$. Here $N'$ is the number of non-zero rows of the matrix $T$. In general $N'\leq N$ because the projector $P_A$ might kill some of the rows. The solutions to Eq. \eqref{eq:deg_dark_matrix} form themselves a vector space of dimension $N' - \rank(T)$. Let us indicate by $\{\vec{a}_i\}_{i=1}^{N'-\rank(T)}$ a basis of this vector space of solutions, i.e., of $\ker(T)$. Once such a basis has been found, it is enough to orthonormalize it using the Gram-Schmidt procedure giving a new basis $\{\vec{c}_i\}_{i=1}^{N'-\rank(T)}$. In this way we will produce a basis of orthonormal eigenstates for the degenerate eigenspace because if $\ket{\psi} = \sum_{m=1}^g c_m\ket{\mu_m}$ and $\ket{\psi'} = \sum_{m=1}^g c_m' \ket{\mu_m}$ are two such states belonging to the eigenspace
\begin{equation}
    \braket{\psi|\psi'} = \vec{c}\cdot \vec{c}'=\delta_{\vec{c},\vec{c}'}\,.
\end{equation}

Repeating the procedure for all degenerate eigenspaces we will eventually find all the dark states satisfying the definition Eq. \eqref{eq:dark_subspace2}.

By the above construction if $N'-\rank(T) \geq 1$ the system admits solutions and dark states exist. Then, according to the discussion in Section \ref{section:dark_states} we have that the probability to eventually detect the target is $P_{\rm det}<1$  also implying an infinite mean first detection time when starting the evolution from one of these states.

We note that these 'infinite hitting` times are well known in the literature for stroboscopic type protocols. For instance, in the context of 'coined` quantum walks on graphs \cite{krovi_2006} it is found that whenever the time evolution operator has sufficient degeneracies the dark subspace will be non trivial. For the same setup, in \cite{Thiel_2020} an upper bound to the total detection probability (the probability to eventually find the target) is derived in terms of the number 'physically equivalent` initial states, i.e., states that yeld the same transition amplitude when measuring the target.

\section{Diagonalization of the Hamiltonian}\label{appendix:nice_matrix}
Consider the matrix
\begin{equation}
    A = \begin{pmatrix}
        a & b & \dots & b\\
        b & a & \dots & b\\
        b & b &\dots & b\\
        \vdots & \vdots & \vdots & \vdots\\
        b & b & \dots & a
    \end{pmatrix}
\end{equation}
with the diagonal elements equal to $a$ and off diagonals equal to $b$. 

Using Gauss elimination one easily finds
\begin{equation}
    \det A = [a + (N-1)b](a-b)^{N-1}\quad .
\end{equation}
Setting $a=1-\Lambda$ and $b=1$ we obtain the characteristic equation for the matrix $E$ with all $1$
\begin{equation}
    (N- \Lambda )\Lambda^{N-1}=0
\end{equation}
which tells that the eigenvalues are $\Lambda =0$ and $\Lambda = N$. For $\Lambda=0$ we need $N-1$ eigenvectors for the matrix to be diagonalizable. We can label the eigenstates by two indices, one referring to the eigenvalue, the second to the degeneracy. That is
\begin{equation}
    \ket{\Lambda, l}\quad,\quad  l=1,2,\dots g_\Lambda
\end{equation}
are the eigenstates associated to the eigenvalue $\Lambda$ with degeneracy $g_\Lambda$ and $l$ is the index labeling the degeneracy. The normalized eigenstate associated to $\Lambda=N$ is not degenerate and is given by
\begin{equation}\label{eq:lambda_non_degenerate}
    \ket{\Lambda=N} = \frac{1}{\sqrt{N}}\begin{pmatrix}
        1 \\ \vdots \\1
    \end{pmatrix}
\end{equation}
while those associated to $\Lambda = 0$ are 
\begin{equation}
    \ket{\Lambda = 0,l=1}=\begin{pmatrix}
        -1 \\ 1 \\ 0 \\ \vdots \\0 
    \end{pmatrix}\quad , \quad \ket{\Lambda = 0, l=2}=\begin{pmatrix}
        -1 \\ 0 \\ 1 \\ \vdots \\0 
    \end{pmatrix}\quad \dots \quad
\end{equation}
and so on. Note that these are not orthogonal between them. We can see by direct inspection that a set of orthonormal $N-1$ eigenstates corresponding to the eigenvalue $\Lambda =0$ is provided by vectors with components 
\begin{equation}\label{eq:orthogonalised_overlap}
    \braket{x|\Lambda=0,l} = -\frac{1}{\sqrt{l(l+1)}}\sum_{i=1}^l \delta_{xi} + \delta_{xl+1}\sqrt{\frac{l}{l+1}}\quad .
\end{equation}
For example
\begin{equation}
    \ket{\Lambda=0, 1}= \begin{pmatrix}
        -\frac{1}{\sqrt{2}}\\
        \frac{1}{\sqrt{2}}
        \\
        0
        \\
        \vdots
        \\
        0
    \end{pmatrix}
    \quad , \quad
    \ket{\Lambda=0, 2}= \begin{pmatrix}
        -\frac{1}{\sqrt{6}}\\
        -\frac{1}{\sqrt{6}}
        \\
        \sqrt{\frac{2}{3}}
        \\
        0
        \\
        \vdots
        \\
        0
    \end{pmatrix}
\end{equation}
and so on.

\section{Calculation of scalar products}\label{app:scalar_products}
In this appendix we report the calculation of the scalar products appearing in Eq. \eqref{eq:laplace_transform_survival}. 
We recall that (see Eq. \eqref{eq:target_subspace})
\begin{equation}
    A = [m+1,N] \quad\text{and } \quad A^\perp \equiv [1,m] 
\end{equation}
are the target subspace and its complement, respectively.

There are two basic scalar products that we need and they are
\begin{equation}\label{eq:scalar_prod_1}
    \braket{N|\psi_0}
\end{equation}
\begin{equation}\label{eq:scalar_prod_2}
    \sum_{l=m}^{N-1}C_l \braket{0,l|\psi_0}\,.
\end{equation}
First of all we split the state as
\begin{equation}
    \ket{\psi_0} = P_A\ket{\psi_0} + P_{A^{\perp}}\ket{\psi_0}
\end{equation}
separating the part in $A$ and the part in $A^\perp$. Since the scalar product is linear in both arguments, we can focus on each component, in $A$ and $A^\perp$ separately and sum the result at the end. 

The easiest one is
\begin{equation}\label{eq:tot_sum_rule1}
    \braket{N|\psi_0} = \frac{1}{\sqrt{N}}\sum_{x\in A^\perp}\psi_0(x) + \frac{1}{\sqrt{N}}\sum_{x\in A}\psi_0(x) = \frac{c_{A^{\perp}} + c_A}{\sqrt{N}}
\end{equation}
where $c_A$ and $c_A^\perp$ are the sum of the coordinates of the initial state in $A$ and $A^\perp$ as defined in the main text in Eq. \eqref{eq:cAdef}.

The product in Eq. \eqref{eq:scalar_prod_2} is slightly more involved. We first consider the first part contributing to Eq. \eqref{eq:scalar_prod_2}
\begin{align}
    \braket{0,l|P_A|\psi_0} &= \sum_{x\in A} \braket{0,l|x}\psi_0(x)\quad , \quad l=m,\dots, N-1\quad , \quad A=[m+1,N]
\end{align}
and recalling the explicit expression of the eigenstates from  Eq. \eqref{eq:normalized_degenerate_states} in the main text,  since $\psi_0(x) = 0$ for $x \leq m$, we obtain that 
\begin{align}
    \braket{0,m|P_A|\psi_0} &= C_m(-\psi_0(m) + m\psi_0(m+1))= C_m m\psi_0(m+1)
    \\
    \braket{0,m+1|P_A|\psi_0} &= C_{m+1}(-\psi_0(m) -\psi_0(m+1) + (m+2) \psi_0(m+2))
\end{align}
where $C_m$ is also defined in Eq. \eqref{eq:normalized_degenerate_states} in the main text. The pattern is then clear and indeed
\begin{equation}
    \braket{0,l|P_A|\psi_0} = C_l\left(-\sum_{x=m}^{l}\psi_0(x) + l\psi_0(l+1)\right)\quad l=m,\dots\,N-1\quad .
\end{equation}
Multiplying by $C_l$ and summing we get
\begin{align}\label{eq:first_sum}
    \sum_{l=m}^{N-1}C_l\braket{0,l|P_A|\psi_0} &=   \sum_{l=m}^{N-1}|C_l|^2\left(-\sum_{x=m}^{l}\psi_0(x) + l\psi_0(l+1)\right)
    \nonumber
    \\
    & = \sum_{l=m}^{N-1}\frac{1}{l(l+1)}\left(-\sum_{x=m}^{l}\psi_0(x) + l\psi_0(l+1)\right)
    \nonumber
    \\
    & = -\sum_{l=m}^{N-1}\frac{1}{l(l+1)}\sum_{x=m}^{l}\psi_0(x) +\sum_{l=m}^{N-1}\frac{1}{l+1}\psi_0(l+1)\quad .
\end{align}
Defining momentarily $A_l= \sum_{x=m}^{l}\psi_0(x)$ we deal with the second term on the right hand side of the last line 
\begin{align}
    \sum_{l=m}^{N-1}\frac{1}{l(l+1)}A_l& = \sum_{l=m}^{N-1}\left(\frac{1}{l}-\frac{1}{l+1}\right)A_l 
    \nonumber
    \\
    & = \frac{A_m}{m} - \frac{A_m}{m+1} + \frac{A_{m+1}}{m+1} - \frac{A_{m+1}}{m+2} + \dots
    \nonumber
    \\
    & = \frac{A_m}{m} + \frac{1}{m+1}\left(A_{m+1} - A_{m}\right) + \dots
    \nonumber
    \\
    & = \frac{A_{m}}{m}+\sum_{l=m+1}^{N}\frac{1}{l}\left(A_{l} - A_{l-1}\right) -A_N/N\quad .
\end{align}
Noting that
\begin{equation}
    A_l - A_{l-1} = \sum_{x=m}^{l} \psi_0(x)-\sum_{x=m}^{l-1} \psi_0(x) = \psi_0(l)
\end{equation}
we obtain
\begin{equation}
    \sum_{l=m}^{N-1}\frac{1}{l(l+1)}\sum_{x=m}^{l}\psi_0(x) =\psi_0(m)/m +\sum_{l=m+1}^{N}\frac{\psi_0(l)}{l} -\frac{1}{N}\sum_{l=m}^N \psi_0(l)\quad .
\end{equation}
This result, plugged in the last line of Eq. \eqref{eq:first_sum} gives 
\begin{align}\label{eq:first_sum_rule2}
    \sum_{l=m}^{N-1}C_l\braket{0,l|P_A|\psi_0} &=-\sum_{l=m}^{N-1}\frac{1}{l(l+1)}\sum_{x=m}^{l}\psi_0(x) +\sum_{l=m}^{N-1}\frac{1}{l+1}\psi_0(l+1) 
    \nonumber
    \\
    &=-\psi_0(m)/m -\sum_{l=m+1}^{N}\frac{\psi_0(l)}{l} +\frac{1}{N}\sum_{l=m}^N \psi_0(l)  +\sum_{l=m}^{N-1}\frac{1}{l+1}\psi_0(l+1)
    \nonumber
    \\
    &=\frac{1}{N}\sum_{l=m}^N \psi_0(l) = \frac{1}{N}\sum_{l=m+1}^N \psi_0(l)
\end{align}
where we used that $\psi_0(m)=0$ as well and
in the last line we changed variables in $l\to l-1$ in the last sum of the second line and simplified with the first one. Thus, given that $A=[m+1,N]$ this is just
\begin{equation}\label{eq:c0_definition}
    c_A =\sum_{l=m+1}^N \psi_0(l)\,.
\end{equation}
We can then rewrite the sum rule Eq. \eqref{eq:first_sum_rule2} as
\begin{equation}\label{eq:sum_rule}
    \sum_{l=m}^{N-1}C_l \braket{0,l|P_A|\psi_0} = \frac{c_A}{N}\quad .
\end{equation}

Then, from \eqref{eq:normalized_degenerate_states}
 in the main text, we see that the first $l$ coordinates of $\ket{0,l}$ are equal $-C_l$ while the $l+1$-th is $l C_l$ and that $\psi_0(x)=0$ for $x\geq m+1$ because the state is localized in $A^\perp$. With these observations we can write
\begin{align}
    \braket{0,l|P_{A^\perp}|\psi_0} = -C_l\sum_{x=1}^{m}\psi_0(x)=-C_l c_{A^{\perp}}\quad ,\quad l=m,\dots, N-1
\end{align}
from which
\begin{equation}\label{eq:sum_rule2}
    \sum_{l=m}^{N-1}C_l\braket{0,l|P_{A^\perp}|\psi_0}=-c_{A^{\perp}}\frac{N-m}{Nm}
\end{equation}
where we have used $\sum_{l=m}^{N-1}C_l^2 =\sum_{l=m}^{N-1}\frac{1}{l(l+1)} = \frac{N-m}{Nm}$. 

This result together with Eq. \eqref{eq:sum_rule} gives the result for Eq. \eqref{eq:scalar_prod_2}
\begin{equation}\label{eq:tot_sum_rule2}
     \sum_{l=m}^{N-1}C_l\braket{0,l|\psi_0} =\frac{c_A}{N} -c_{A^{\perp}}\frac{N-m}{Nm}\,.
\end{equation}

\section{Routh-Hurwitz criterion and decay of $F(t)$}\label{app:hurwitz}

Consider the cubic polynomial with real coefficients
\begin{equation}\label{eq:generic}
Q(s)=s^{3}+a_{2}s^{2}+a_{1}s+a_{0}, \qquad a_{k}\in\mathbb R.
\end{equation}
The \emph{Routh-Hurwitz criterion} for a third degree polynomial with unit leading coefficient says that \emph{all} the zeros of \eqref{eq:generic} satisfy $\mathrm{Re}\,s<0$ if and only the following Routh-Hurwitz inequalities hold \cite{Weisstein_RouthHurwitz, Wikipedia_RouthHurwitz}
\begin{equation}\label{eq:RHcriterion}
 a_{2}>0, \quad a_{1}>0, \quad a_{0}>0, \quad a_{2}a_{1}-a_{0}>0.
\end{equation}
For a cubic these four inequalities are both necessary and sufficient for having all the roots strictly on the left half complex plane.
Let us consider the polynomial in Eq. \eqref{eq:poly_den}
\begin{equation}\label{eq:poly_den2}
Q(s)=s^{3}+2r\,s^{2}+\bigl(J^{2}N^{2}+r^{2}\bigr)s+2J^{2}mr\,(N-m),
\end{equation}
with
\begin{equation}\label{eq:params}
 r>0, \qquad J\neq0, \qquad N\ge1, \qquad 0<m\le N.
\end{equation}
Comparing \eqref{eq:poly_den2} with \eqref{eq:generic} we read off
\begin{equation}\label{eq:coeffs}
 a_{2}=2r,\qquad a_{1}=J^{2}N^{2}+r^{2},\qquad a_{0}=2J^{2}mr\,(N-m).
\end{equation}
The first three Routh-Hurwitz inequalities in \eqref{eq:RHcriterion} are immediate from \eqref{eq:params} and \eqref{eq:coeffs}.  The last one reads
\begin{equation}\label{eq:RHdet}
 a_{2}a_{1}-a_{0}=2r\bigl(J^{2}N^{2}+r^{2}\bigr)-2J^{2}mr\,(N-m)
               =2r\Bigl[J^{2}\bigl(N^{2}-m(N-m)\bigr)+r^{2}\Bigr]>0,
\end{equation}
where we used $N^{2}-m(N-m)=m^{2}+(N-m)^{2}\ge0$.  Hence \eqref{eq:RHcriterion} holds and every zero of $Q(s)$ lies strictly in the open left half-plane. 

Next we show that the real negative root $s_1$ determines the slowest decay rate, i.e., $s_R<s_1<0$ where $s_R = \Re s_{2,3}$ denotes the imaginary part of the two conjugate roots. Looking at Eq. \eqref{eq:exact_F} we need to show what $r>|s_1|$ and $|s_R|>|s_1|$. From the Routh-Hurwitz criterion we know that
\begin{equation}\label{eq:roots}
 s_{1}<0,\qquad s_{2}=s_{R}+i s_{I},\qquad s_{3}=s_{R}-i s_{I}\quad, \quad s_{I}\neq 0\quad ,\quad s_R<0\,.
\end{equation}
To bound the roots, we look for a sign change of $Q(s)$ on the negative real axis and use the Intermediate Value Theorem which says that for a continuous function if $Q(a)Q(b)<0$ then there is a zero in $[a,b]$.
Evaluating $Q(s)$ at $s=0$, from \eqref{eq:poly_den2}
\begin{equation}\label{eq:Q0again}
    Q(0)=2J^{2}mr\,(N-m)>0\,.
\end{equation}
Next we evaluate $Q(s)$ at $s=-\frac{2}{3}r$
\begin{equation}\label{eq:Qminus23r_prel}
 Q\bigl(-\frac{2}{3} r\bigr)=2r\Bigl[-\frac{1}{3}J^{2}N^{2}-\tfrac1{27}r^{2}+J^{2}m(N-m)\Bigg]\,.
\end{equation}
To show that this is negative we note that
\begin{equation}
    m(N-m) = \frac{N^2}{4} - (m-N/2)^2\leq \frac{N^2}{4}
\end{equation}
that used in the square bracket in Eq. \eqref{eq:Qminus23r_prel} gives
\begin{equation}\label{eq:Qminus23r}
    Q\bigl(-\frac{2}{3} r\bigr)\leq 2r\Bigl[-\frac{1}{12}J^{2}N^{2}-\tfrac1{27}r^{2}\Bigg]< 0\,.
\end{equation}
Combining \eqref{eq:Q0again} and \eqref{eq:Qminus23r} the Intermediate Value Theorem implies that
\begin{equation}\label{eq:s1tighter}
 -\frac{2}{3} r<s_{1}<0\,.
\end{equation}
In particular, since $r>0$
\begin{equation}\label{eq:rlesss1}
    -r<s_1<0\,.
\end{equation}
Then we use the following sum rule satisfied by the roots of the a generic third order polynomial as in Eq. \eqref{eq:generic}
\begin{equation}
    s_1 + s_2 + s_3 = a_2
\end{equation}
which applied to $Q(s)$ in Eq. \eqref{eq:poly_den2}, using Eq. \eqref{eq:roots}, translates into
\begin{equation}\label{eq:Viete}
 s_{1}+2s_{R}=-2r \,.
\end{equation}
Using the fact that $s_1<0$ and $r>0$ we get
\begin{equation}
     s_{R}=-\frac{2r+s_{1}}{2}<0\,.
\end{equation}
Subtracting from $s_1$ and using Eq. \eqref{eq:rlesss1} we obtain
\begin{equation}\label{eq:difference}
 s_{1}-s_{R}=\frac{3s_{1}+2r}{2}>0\,.
\end{equation}
Hence
\begin{equation}\label{eq:sRlesss1}
s_{R}<s_{1}<0\,,
\end{equation}
meaning the complex conjugate pair lies further left than the real root.

Together \eqref{eq:sRlesss1} and Eq. \eqref{eq:rlesss1} imply $|s_{R}|>|s_{1}|$ and $r>|s_1|$ respectively. This means that the term $e^{s_{1}t}$ in Eq. \eqref{eq:exact_F} sets the longest decay time
\begin{equation}\label{eq:tmax}
 t_{m}(r)=\frac{1}{|s_{1}|}\,,
\end{equation}
while the exponential $e^{-rt}$ and the oscillatory contributions from $s_{2,3}$ are suppressed faster.

\section{The special state $\ket{\psi^*}$}\label{app:special}
Here we prove that the special state in Eq. \eqref{eq:special_state} is the unique bright state with the property of being \emph{completely} localized in $A^\perp$, i.e., $P_A\ket{\psi^*}=0$.  The facts that the state is bright and completely localized in $A$ follow from its definition. Indeed, using the property $P_D P_{A^\perp} = P_D$ (see Section \ref{section:dark_states}), we can see that that $P_D \ket{\psi^*} \propto P_D \ket{N}=0$ because $\ket{N}$ is a bright eigenstate (it overlaps with the measured region). This means that $\ket{\psi^*}$ is a bright state despite not having components in $A$. Are there other states that have no component in $A$ but that are bright ? The answer is no. An argument appealing to physical intuition is that if such states were to exist we would have found other states for which the first detection probability $F(t)\sim t^a$ for $t\to 0$ with some exponent $a$. Since from Eq. \eqref{eq:F_hat_lapl_small_s} this cannot happen we are led to conclude that these states do not exist. To see this more concretely  one considers a generic bright state $\ket{\phi} = \sum_{l=1}^{N-m-1}a_l \ket{0,l+m} + a_{N-m}\ket{N}$ and solves the linear system
\begin{equation}
    P_A \ket{\phi}=\sum_{l=1}^{N-m}a_l\, P_A\ket{0,l+m-1} + a_{N-m+1}P_A\ket{N}=0  
\end{equation}
for the coefficients $a_l$. In matrix form this is 
\begin{equation}
    \begin{pmatrix}
        \vline & \vline &  & \vline & \vline \\
        P_A \ket{0,m} & P_A \ket{0,m+1} & \cdots & P_A \ket{0,N-1} & P_A \ket{N}\\
        \vline & \vline &  & \vline & \vline 
        \end{pmatrix}\begin{pmatrix}
        a_{1} \\ \vdots \\a_{N-m+1} 
    \end{pmatrix}=0\,.
\end{equation}
The solutions to this $N\times N-m+1$ system provide all the bright states that have no component in $A$. Recalling the particular form of the eigenstates $\ket{0,l}$ in Eq. \eqref{eq:normalized_degenerate_states}, the fact that $P_A$ kills the first $m$ coordinates of a vector leads to

\begin{equation}\left(
\begin{array}{cccccccc}
0 & 0 & 0 & \cdots & 0 & 0 & 0 \\
0 & 0 & 0 & \cdots & 0 & 0 & 0 \\
\vdots & \vdots & \vdots &  & \vdots & \vdots & \vdots \\
0 & 0 & 0 & \cdots & 0 & 0 & 0 \\[8pt]
m   & -1 & -1 & \cdots & -1 & -1 & 1\\
0   & m+1 & -1 & \cdots & -1 & -1 & 1\\
0   & 0   & m+2 & \ddots & \vdots & \vdots & 1\\
\vdots &  &  & \ddots & -1 & -1 & 1\\
0 & \cdots & 0 & 0 & N-2 & -1 & 1\\
0 & \cdots & 0 & 0 & 0 & N-1 & 1
\end{array}
\right)
\begin{pmatrix}
a_1\\
\vdots\\
a_{N-m+1}
\end{pmatrix}
\;=\;0\,.
\end{equation}
The full matrix on the left has dimension $N \times N-m+1$ while the lower submatrix has dimension $N-m\times N-m+1$. It is easy to see that the rank of this submatrix is $N-m$ as all the rows a linearly independent. This implies that also the rank of the full matrix is $N-m$. As a consequence the number of linearly independent non-zero solution for the vector of coefficients is given by (by the rank-nullity theorem in linear algebra)
\begin{equation}
    N-m+1 - (N-m) = 1\,.
\end{equation}
This solution corresponds exactly to the state $\ket{\psi^*}$ since we know that it is both bright and it does not have components in $A$, i.e., $P_A\ket{\psi^*}=0$ (see Eq. \eqref{eq:special_state}).

\end{appendices}

\section*{Acknowledgements}

We acknowledge support from ANR Grant No. ANR-23-CE30-0020-01 EDIPS.

\printbibliography

@article{Kulkarni_2023,
doi = {10.1088/1751-8121/acf103},
url = {https://dx.doi.org/10.1088/1751-8121/acf103},
year = {2023},
month = {sep},
publisher = {IOP Publishing},
volume = {56},
number = {38},
pages = {385003},
author = {Manas Kulkarni and Satya N Majumdar},
title = {First detection probability in quantum resetting via random projective measurements},
journal = {J. Phys. A: Math. Theor.},
}

@article{Nagar_2023,
doi = {10.1088/1751-8121/acda6c},
url = {https://dx.doi.org/10.1088/1751-8121/acda6c},
year = {2023},
month = {jun},
publisher = {IOP Publishing},
volume = {56},
number = {28},
pages = {283001},
author = {Nagar, Apoorva and Gupta, Shamik},
title = {Stochastic resetting in interacting particle systems: a review},
journal = {J. Phys. A: Math. Theor.}
}

@article{Pal_2022,
doi = {10.1088/1751-8121/ac3cdf},
url = {https://dx.doi.org/10.1088/1751-8121/ac3cdf},
year = {2022},
month = {jan},
publisher = {IOP Publishing},
volume = {55},
number = {2},
pages = {021001},
author = {Pal, Arnab and Kostinski, Sarah and Reuveni, Shlomi},
title = {The inspection paradox in stochastic resetting},
journal = {J. Phys. A: Math. Theor.}
}

@article{Thiel_2020,
  title = {Dark states of quantum search cause imperfect detection},
  author = {Thiel, Felix and Mualem, Itay and Meidan, Dror and Barkai, Eli and Kessler, David A.},
  journal = {Phys. Rev. Res.},
  volume = {2},
  issue = {4},
  pages = {043107},
  numpages = {16},
  year = {2020},
  month = {Oct},
  publisher = {American Physical Society},
  doi = {10.1103/PhysRevResearch.2.043107},
  url = {https://link.aps.org/doi/10.1103/PhysRevResearch.2.043107}
}

@article{yin2019,
  title = {Large fluctuations of the first detected quantum return time},
  author = {Yin, R. and Ziegler, K. and Thiel, F. and Barkai, E.},
  journal = {Phys. Rev. Res.},
  volume = {1},
  issue = {3},
  pages = {033086},
  numpages = {14},
  year = {2019},
  month = {Nov},
  publisher = {American Physical Society},
  doi = {10.1103/PhysRevResearch.1.033086},
  url = {https://link.aps.org/doi/10.1103/PhysRevResearch.1.033086}
}

@article{krovi_2006,
  title = {Quantum walks with infinite hitting times},
  author = {Krovi, Hari and Brun, Todd A.},
  journal = {Phys. Rev. A},
  volume = {74},
  issue = {4},
  pages = {042334},
  numpages = {11},
  year = {2006},
  month = {Oct},
  publisher = {American Physical Society},
  doi = {10.1103/krovi_2006},
  url = {https://link.aps.org/doi/10.1103/krovi_2006}
}

@article{verbanov_2008,
  title = {Hitting time for the continuous quantum walk},
  author = {Varbanov, Martin and Krovi, Hari and Brun, Todd A.},
  journal = {Phys. Rev. A},
  volume = {78},
  issue = {2},
  pages = {022324},
  numpages = {12},
  year = {2008},
  month = {Aug},
  publisher = {American Physical Society},
  doi = {10.1103/PhysRevA.78.022324},
  url = {https://link.aps.org/doi/10.1103/PhysRevA.78.022324}
}

@incollection{GDM2024,
  title={Target Search Problems},
  author={Grebenkov, Denis and Metzler, Ralf and Oshanin, Gleb},
  booktitle={Target Search Problems},
  pages={},
  year={2024},
  publisher={Springer}
}

@Article{Grunbaum2013,
author={Gr{\"u}nbaum, F. A.
and Vel{\'a}zquez, L.
and Werner, A. H.
and Werner, R. F.},
title={Recurrence for Discrete Time Unitary Evolutions},
journal={Commun.\ Math.\ Phys.},
year={2013},
month={Jun},
day={01},
volume={320},
number={2},
pages={543-569},
issn={1432-0916},
doi={10.1007/s00220-012-1645-2},
url={https://doi.org/10.1007/s00220-012-1645-2}
}

@article{Kessler_2021,
  title = {First-detection time of a quantum state under random probing},
  author = {Kessler, David A. and Barkai, Eli and Ziegler, Klaus},
  journal = {Phys. Rev. A},
  volume = {103},
  issue = {2},
  pages = {022222},
  numpages = {10},
  year = {2021},
  month = {Feb},
  publisher = {American Physical Society},
  doi = {10.1103/PhysRevA.103.022222},
  url = {https://link.aps.org/doi/10.1103/PhysRevA.103.022222}
}

@article{Friedman_2017,
doi = {10.1088/1751-8121/aa5191},
url = {https://dx.doi.org/10.1088/1751-8121/aa5191},
year = {2016},
month = {dec},
publisher = {IOP Publishing},
volume = {50},
number = {4},
pages = {04LT01},
author = {H Friedman and D A Kessler and E Barkai},
title = {Quantum renewal equation for the first detection time of a quantum walk},
journal = {J. Phys. A: Math. Theor.}
}

@article{Perfetto2021,
  title = {Designing nonequilibrium states of quantum matter through stochastic resetting},
  author = {Perfetto, Gabriele and Carollo, Federico and Magoni, Matteo and Lesanovsky, Igor},
  journal = {Phys. Rev. B},
  volume = {104},
  issue = {18},
  pages = {L180302},
  numpages = {6},
  year = {2021},
  month = {Nov},
  publisher = {American Physical Society},
  doi = {10.1103/PhysRevB.104.L180302},
  url = {https://link.aps.org/doi/10.1103/PhysRevB.104.L180302}
}

@article{kulkarni_2023b,
  title = {Generating entanglement by quantum resetting},
  author = {Kulkarni, Manas and Majumdar, Satya N.},
  journal = {Phys. Rev. A},
  volume = {108},
  issue = {6},
  pages = {062210},
  numpages = {13},
  year = {2023},
  month = {Dec},
  publisher = {American Physical Society},
  doi = {10.1103/PhysRevA.108.062210},
  url = {https://link.aps.org/doi/10.1103/PhysRevA.108.062210}
}

@Article{wang2024hitting,
AUTHOR = {Wang, Qingyuan and Ren, Silin and Yin, Ruoyu and Ziegler, Klaus and Barkai, Eli and Tornow, Sabine},
TITLE = {First Hitting Times on a Quantum Computer: Tracking vs. Local Monitoring, Topological Effects, and Dark States},
JOURNAL = {Entropy},
VOLUME = {26},
YEAR = {2024},
NUMBER = {10},
ARTICLE-NUMBER = {869},
URL = {https://www.mdpi.com/1099-4300/26/10/869},
PubMedID = {39451946},
ISSN = {1099-4300},
DOI = {10.3390/e26100869}
}

@book{redner2001,
  author    = {S. Redner},
  title     = {A Guide to First-Passage Processes},
  publisher = {Cambridge University Press},
  year      = {2001}
}

@book{metzler2014,
  author    = {R. Metzler and G. Oshanin and S. Redner},
  title     = {First-Passage Phenomena and Their Applications},
  publisher = {World Scientific},
  year      = {2014}
}

@article{MAEG09,
doi = {10.1088/1751-8121/42/43/430301},
url = {https://dx.doi.org/10.1088/1751-8121/42/43/430301},
year = {2009},
month = {oct},
publisher = {},
volume = {42},
number = {43},
pages = {430301},
author = {Marcos G E da Luz and Alexander Grosberg and Ernesto P Raposo and Gandhi M Viswanathan},
title = {The random search problem: trends and perspectives},
journal = {J. Phys. A: Math. Theor.},
}

@online{Weisstein_RouthHurwitz,
  author={Weisstein, Eric W.},
  title= {Routh--Hurwitz Theorem},
  year         = {2004},           
  organization = {Wolfram Research},
  howpublished = {\\textit{MathWorld} -- A Wolfram Web Resource},
  url          = {https://mathworld.wolfram.com/Routh-HurwitzTheorem.html},
  note         = {Accessed: 2025-07-10}
}

@online{Wikipedia_RouthHurwitz,
  author       = {{Wikipedia contributors}},
  title        = {Routh--Hurwitz Stability Criterion},
  year         = {2024},             
  howpublished = {\emph{Wikipedia, The Free Encyclopedia}},
  url          = {https://en.wikipedia.org/wiki/Routh%E2%80%93Hurwitz_stability_criterion},
  note         = {Accessed: 2025-07-10}
}

@Inbook{F2015,
author="Andrad{\'o}ttir, Sigr{\'u}n",
editor="Fu, Michael C",
title="A Review of Random Search Methods",
bookTitle="Handbook of Simulation Optimization",
year="2015",
publisher="Springer New York",
address="New York, NY",
isbn="978-1-4939-1384-8",
doi="10.1007/978-1-4939-1384-8_10",
url="https://doi.org/10.1007/978-1-4939-1384-8_10"
}

@book{sakurai2020,
  author    = {J. J. Sakurai and J. Napolitano},
  title     = {Modern Quantum Mechanics},
  edition   = {3rd},
  publisher = {Cambridge University Press},
  year      = {2020},
  address   = {Cambridge}
}

@article{misra1977,
    author = {Misra, B. and Sudarshan, E. C. G.},
    title = {The Zeno’s paradox in quantum theory},
    journal = {Journal of Mathematical Physics},
    volume = {18},
    number = {4},
    pages = {756-763},
    year = {1977},
    month = {04},
    issn = {0022-2488},
    doi = {10.1063/1.523304},
    url = {https://doi.org/10.1063/1.523304},
    eprint = {https://pubs.aip.org/aip/jmp/article-pdf/18/4/756/19182345/756_1_online.pdf},
}

@article{itano1990,
  title = {Quantum Zeno effect},
  author = {Itano, Wayne M. and Heinzen, D. J. and Bollinger, J. J. and Wineland, D. J.},
  journal = {Phys. Rev. A},
  volume = {41},
  issue = {5},
  pages = {2295--2300},
  numpages = {0},
  year = {1990},
  month = {Mar},
  publisher = {American Physical Society},
  doi = {10.1103/PhysRevA.41.2295},
  url = {https://link.aps.org/doi/10.1103/PhysRevA.41.2295}
}

@article{Dhar_2015b,
  title = {Detection of a quantum particle on a lattice under repeated projective measurements},
  author = {Dhar, Shrabanti and Dasgupta, Subinay and Dhar, Abhishek and Sen, Diptiman},
  journal = {Phys. Rev. A},
  volume = {91},
  issue = {6},
  pages = {062115},
  numpages = {10},
  year = {2015},
  month = {Jun},
  publisher = {American Physical Society},
  doi = {10.1103/PhysRevA.91.062115},
  url = {https://link.aps.org/doi/10.1103/PhysRevA.91.062115}
}

@article{childs_2014,
  title = {Spatial search by continuous-time quantum walks on crystal lattices},
  author = {Childs, Andrew M. and Ge, Yimin},
  journal = {Phys. Rev. A},
  volume = {89},
  issue = {5},
  pages = {052337},
  numpages = {11},
  year = {2014},
  month = {May},
  publisher = {American Physical Society},
  doi = {10.1103/PhysRevA.89.052337},
  url = {https://link.aps.org/doi/10.1103/PhysRevA.89.052337}
}

@article{farhi_1998,
  title = {Quantum computation and decision trees},
  author = {Farhi, Edward and Gutmann, Sam},
  journal = {Phys. Rev. A},
  volume = {58},
  issue = {2},
  pages = {915--928},
  numpages = {0},
  year = {1998},
  month = {Aug},
  publisher = {American Physical Society},
  doi = {10.1103/PhysRevA.58.915},
  url = {https://link.aps.org/doi/10.1103/PhysRevA.58.915}
}

@article{mukherjee2018,
  title = {Quantum dynamics with stochastic reset},
  author = {Mukherjee, B. and Sengupta, K. and Majumdar, Satya N.},
  journal = {Phys. Rev. B},
  volume = {98},
  issue = {10},
  pages = {104309},
  numpages = {14},
  year = {2018},
  month = {Sep},
  publisher = {American Physical Society},
  doi = {10.1103/PhysRevB.98.104309},
  url = {https://link.aps.org/doi/10.1103/PhysRevB.98.104309}
}

@article{Dattagupta_2022,
doi = {10.1088/1742-5468/ac98c0},
url = {https://dx.doi.org/10.1088/1742-5468/ac98c0},
year = {2022},
month = {oct},
publisher = {IOP Publishing and SISSA},
volume = {2022},
number = {10},
pages = {103210},
author = {Dattagupta, Sushanta and Das, Debraj and Gupta, Shamik},
title = {Stochastic resets in the context of a tight-binding chain driven by an oscillating field},
journal = {J. Stat. Mech.: Theory Exp.}
}

@article{Das_2022b,
doi = {10.1088/1742-5468/ac6256},
url = {https://dx.doi.org/10.1088/1742-5468/ac6256},
year = {2022},
month = {may},
publisher = {IOP Publishing and SISSA},
volume = {2022},
number = {5},
pages = {053101},
author = {Das, Debraj and Dattagupta, Sushanta and Gupta, Shamik},
title = {Quantum unitary evolution interspersed with repeated non-unitary interactions at random times: the method of stochastic Liouville equation, and two examples of interactions in the context of a tight-binding chain},
journal = {J. Stat. Mech.: Theory Exp.}
}

@article{Dubey_2023,
doi = {10.1088/1751-8121/acc290},
url = {https://dx.doi.org/10.1088/1751-8121/acc290},
year = {2023},
month = {mar},
publisher = {IOP Publishing},
volume = {56},
number = {15},
pages = {154001},
author = {Dubey, Varun and Chetrite, Raphael and Dhar, Abhishek},
title = {Quantum resetting in continuous measurement induced dynamics of a qubit},
journal = {J. Phys. A: Math. Theor.}
}

@article{Sevilla_2023,
doi = {10.1088/1751-8121/acb29d},
url = {https://dx.doi.org/10.1088/1751-8121/acb29d},
year = {2023},
month = {feb},
publisher = {IOP Publishing},
volume = {56},
number = {3},
pages = {034001},
author = {Sevilla, Francisco J and Valdés-Hernández, Andrea},
title = {Dynamics of closed quantum systems under stochastic resetting},
journal = {J. Phys. A: Math. Theor.}
}

@article{yin_2023,
  title = {Restart Expedites Quantum Walk Hitting Times},
  author = {Yin, R. and Barkai, E.},
  journal = {Phys. Rev. Lett.},
  volume = {130},
  issue = {5},
  pages = {050802},
  numpages = {6},
  year = {2023},
  month = {Feb},
  publisher = {American Physical Society},
  doi = {10.1103/PhysRevLett.130.050802},
  url = {https://link.aps.org/doi/10.1103/PhysRevLett.130.050802}
}

@article{yin_2024,
  title = {Instability in the quantum restart problem},
  author = {Yin, Ruoyu and Wang, Qingyuan and Barkai, Eli},
  journal = {Phys. Rev. E},
  volume = {109},
  issue = {6},
  pages = {064150},
  numpages = {15},
  year = {2024},
  month = {Jun},
  publisher = {American Physical Society},
  doi = {10.1103/PhysRevE.109.064150},
  url = {https://link.aps.org/doi/10.1103/PhysRevE.109.064150}
}

@article{rose2018,
  title = {Spectral properties of simple classical and quantum reset processes},
  author = {Rose, Dominic C. and Touchette, Hugo and Lesanovsky, Igor and Garrahan, Juan P.},
  journal = {Phys. Rev. E},
  volume = {98},
  issue = {2},
  pages = {022129},
  numpages = {10},
  year = {2018},
  month = {Aug},
  publisher = {American Physical Society},
  doi = {10.1103/PhysRevE.98.022129},
  url = {https://link.aps.org/doi/10.1103/PhysRevE.98.022129}
}

@article{Dhar_2015a,
doi = {10.1088/1751-8113/48/11/115304},
url = {https://dx.doi.org/10.1088/1751-8113/48/11/115304},
year = {2015},
month = {feb},
publisher = {IOP Publishing},
volume = {48},
number = {11},
pages = {115304},
author = {Shrabanti Dhar and Subinay Dasgupta and Abhishek Dhar},
title = {Quantum time of arrival distribution in a simple lattice model},
journal = {J. Phys. A: Math. Theor.}
}

@article{tornow_2023,
  title = {Measurement-induced quantum walks on an IBM quantum computer},
  author = {Tornow, Sabine and Ziegler, Klaus},
  journal = {Phys. Rev. Res.},
  volume = {5},
  issue = {3},
  pages = {033089},
  numpages = {10},
  year = {2023},
  month = {Aug},
  publisher = {American Physical Society},
  doi = {10.1103/PhysRevResearch.5.033089},
  url = {https://link.aps.org/doi/10.1103/PhysRevResearch.5.033089}
}

@article{lahiri_2019,
  title = {Return to the origin problem for a particle on a one-dimensional lattice with quasi-Zeno dynamics},
  author = {Lahiri, Sourabh and Dhar, Abhishek},
  journal = {Phys. Rev. A},
  volume = {99},
  issue = {1},
  pages = {012101},
  numpages = {9},
  year = {2019},
  month = {Jan},
  publisher = {American Physical Society},
  doi = {10.1103/PhysRevA.99.012101},
  url = {https://link.aps.org/doi/10.1103/PhysRevA.99.012101}
}

@article{dubey_2021,
  title = {Quantum dynamics under continuous projective measurements: Non-Hermitian description and the continuum-space limit},
  author = {Dubey, Varun and Bernardin, C\'edric and Dhar, Abhishek},
  journal = {Phys. Rev. A},
  volume = {103},
  issue = {3},
  pages = {032221},
  numpages = {17},
  year = {2021},
  month = {Mar},
  publisher = {American Physical Society},
  doi = {10.1103/PhysRevA.103.032221},
  url = {https://link.aps.org/doi/10.1103/PhysRevA.103.032221}
}

@article{kempe_2005,
	author = {Kempe, Julia},
	date = {2005/10/01},
	doi = {10.1007/s00440-004-0423-2},
	id = {Kempe2005},
	isbn = {1432-2064},
	journal = {Probab.\ Theory Relat.\ Fields},
	number = {2},
	pages = {215--235},
	title = {Discrete Quantum Walks Hit Exponentially Faster},
	url = {https://doi.org/10.1007/s00440-004-0423-2},
	volume = {133},
	year = {2005}}

@article{aharonov_1993,
  title = {Quantum random walks},
  author = {Aharonov, Y. and Davidovich, L. and Zagury, N.},
  journal = {Phys. Rev. A},
  volume = {48},
  issue = {2},
  pages = {1687--1690},
  numpages = {0},
  year = {1993},
  month = {Aug},
  publisher = {American Physical Society},
  doi = {10.1103/PhysRevA.48.1687},
  url = {https://link.aps.org/doi/10.1103/PhysRevA.48.1687}
}

@article{krovi_2006b,
  title = {Hitting time for quantum walks on the hypercube},
  author = {Krovi, Hari and Brun, Todd A.},
  journal = {Phys. Rev. A},
  volume = {73},
  issue = {3},
  pages = {032341},
  numpages = {8},
  year = {2006},
  month = {Mar},
  publisher = {American Physical Society},
  doi = {10.1103/PhysRevA.73.032341},
  url = {https://link.aps.org/doi/10.1103/PhysRevA.73.032341}
}

@article{vanegas,
	author = {Venegas-Andraca, Salvador El{\'\i}as},
	date = {2012/10/01},
	doi = {10.1007/s11128-012-0432-5},
	id = {Venegas-Andraca2012},
	isbn = {1573-1332},
	journal = {Quantum Inf.\ Process.},
	number = {5},
	pages = {1015--1106},
	title = {Quantum walks: a comprehensive review},
	url = {https://doi.org/10.1007/s11128-012-0432-5},
	volume = {11},
	year = {2012}}

@article{evans_2011a,
  title = {Diffusion with Stochastic Resetting},
  author = {Evans, Martin R. and Majumdar, Satya N.},
  journal = {Phys. Rev. Lett.},
  volume = {106},
  issue = {16},
  pages = {160601},
  numpages = {4},
  year = {2011},
  month = {Apr},
  publisher = {American Physical Society},
  doi = {10.1103/PhysRevLett.106.160601},
  url = {https://link.aps.org/doi/10.1103/PhysRevLett.106.160601}
}

@article{Evans_2011b,
doi = {10.1088/1751-8113/44/43/435001},
url = {https://dx.doi.org/10.1088/1751-8113/44/43/435001},
year = {2011},
month = {oct},
publisher = {IOP Publishing},
volume = {44},
number = {43},
pages = {435001},
author = {Evans, Martin R and Majumdar, Satya N},
title = {Diffusion with optimal resetting},
journal = {J. Phys. A: Math. Theor.}
}

@article{Evans_2020,
doi = {10.1088/1751-8121/ab7cfe},
url = {https://dx.doi.org/10.1088/1751-8121/ab7cfe},
year = {2020},
month = {apr},
publisher = {IOP Publishing},
volume = {53},
number = {19},
pages = {193001},
author = {Martin R Evans and Satya N Majumdar and Grégory Schehr},
title = {Stochastic resetting and applications},
journal = {J. Phys. A: Math. Theor.}
}

@article{Kulkarni_2025,
doi = {10.1088/1751-8121/adb6db},
url = {https://dx.doi.org/10.1088/1751-8121/adb6db},
year = {2025},
month = {mar},
publisher = {IOP Publishing},
volume = {58},
number = {10},
pages = {105003},
author = {Kulkarni, Manas and Majumdar, Satya N and Sabhapandit, Sanjib},
title = {Dynamically emergent correlations in bosons via quantum resetting},
journal = {J. Phys. A: Math. Theor.}
}

@inproceedings{grover1996,
author = {Grover, Lov K.},
title = {A fast quantum mechanical algorithm for database search},
year = {1996},
isbn = {0897917855},
publisher = {Association for Computing Machinery},
address = {New York, NY, USA},
url = {https://doi.org/10.1145/237814.237866},
doi = {10.1145/237814.237866},
booktitle = {Proceedings of the Twenty-Eighth Annual ACM Symposium on Theory of Computing},
pages = {212–219},
numpages = {8},
location = {Philadelphia, Pennsylvania, USA},
series = {STOC '96}
}

@Article{Giri2017,
author={Giri, Pulak Ranjan
and Korepin, Vladimir E.},
title={A review on quantum search algorithms},
journal={Quantum Inf.\ Process.},
year={2017},
month={Nov},
day={13},
volume={16},
number={12},
pages={315},
issn={1573-1332},
doi={10.1007/s11128-017-1768-7},
url={https://doi.org/10.1007/s11128-017-1768-7}
}

@article{kempe,
author = {J Kempe},
title = {Quantum random walks: An introductory overview},
journal = {Contemporary Physics},
volume = {44},
number = {4},
pages = {307--327},
year = {2003},
publisher = {Taylor \& Francis},
doi = {10.1080/00107151031000110776},


URL = { 
    
        https://doi.org/10.1080/00107151031000110776
    
    

},
eprint = { 
    
        https://doi.org/10.1080/00107151031000110776
    
    

}

}

@inproceedings{ambainis_2001,
author = {Ambainis, Andris and Bach, Eric and Nayak, Ashwin and Vishwanath, Ashvin and Watrous, John},
title = {One-dimensional quantum walks},
year = {2001},
isbn = {1581133499},
publisher = {Association for Computing Machinery},
address = {New York, NY, USA},
url = {https://doi.org/10.1145/380752.380757},
doi = {10.1145/380752.380757},
booktitle = {Proceedings of the Thirty-Third Annual ACM Symposium on Theory of Computing},
pages = {37–49},
numpages = {13},
location = {Hersonissos, Greece},
series = {STOC '01}
}

@article{Montero2013,
  author = {Montero, M. and Villarroel, J.},
  title = {Monotonous continuous-time random walks with drift and stochastic reset events},
  journal = {Phys. Rev. E},
  volume = {87},
  number = {1},
  pages = {012116},
  year = {2013},
  doi = {10.1103/PhysRevE.87.012116},
  url = {https://link.aps.org/doi/10.1103/PhysRevE.87.012116}
}

@article{Evans2014,
  author = {Evans, M. R. and Majumdar, S. N.},
  title = {Diffusion with resetting in arbitrary spatial dimension},
  journal = {J. Phys. A: Math. Theor.},
  volume = {47},
  number = {28},
  pages = {285001},
  year = {2014},
  doi = {10.1088/1751-8113/47/28/285001},
  url = {https://iopscience.iop.org/article/10.1088/1751-8113/47/28/285001}
}

@article{Gupta2014,
  author = {Gupta, S. and Majumdar, S. N. and Schehr, G.},
  title = {Fluctuating interfaces subject to stochastic resetting},
  journal = {Phys. Rev. Lett.},
  volume = {112},
  number = {22},
  pages = {220601},
  year = {2014},
  doi = {10.1103/PhysRevLett.112.220601},
  url = {https://link.aps.org/doi/10.1103/PhysRevLett.112.220601}
}

@article{Pal2015,
  author = {Pal, A.},
  title = {Diffusion in a potential landscape with stochastic resetting},
  journal = {Phys. Rev. E},
  volume = {91},
  number = {1},
  pages = {012113},
  year = {2015},
  doi = {10.1103/PhysRevE.91.012113},
  url = {https://link.aps.org/doi/10.1103/PhysRevE.91.012113}
}

@article{Majumdar2015,
  author = {Majumdar, S. N. and Sabhapandit, S. and Schehr, G.},
  title = {Dynamical transition in the temporal relaxation of stochastic processes under resetting},
  journal = {Phys. Rev. E},
  volume = {91},
  number = {5},
  pages = {052131},
  year = {2015},
  doi = {10.1103/PhysRevE.91.052131},
  url = {https://link.aps.org/doi/10.1103/PhysRevE.91.052131}
}

@article{Christou2015,
  author = {Christou, C. and Schadschneider, A.},
  title = {Diffusion with resetting in bounded domains},
  journal = {J. Phys. A: Math. Theor.},
  volume = {48},
  number = {28},
  pages = {285003},
  year = {2015},
  doi = {10.1088/1751-8113/48/28/285003},
  url = {https://iopscience.iop.org/article/10.1088/1751-8113/48/28/285003}
}

@article{Montero2016,
  author = {Montero, M. and Villarroel, J.},
  title = {Directed random walk with random restarts: The Sisyphus random walk},
  journal = {Phys. Rev. E},
  volume = {94},
  number = {3},
  pages = {032132},
  year = {2016},
  doi = {10.1103/PhysRevE.94.032132},
  url = {https://link.aps.org/doi/10.1103/PhysRevE.94.032132}
}

@article{Mendez2016,
  author = {Mendez, V. and Campos, D.},
  title = {Characterization of stationary states in random walks with stochastic resetting},
  journal = {Phys. Rev. E},
  volume = {93},
  number = {2},
  pages = {022106},
  year = {2016},
  doi = {10.1103/PhysRevE.93.022106},
  url = {https://link.aps.org/doi/10.1103/PhysRevE.93.022106}
}

@article{Eule2016,
  author = {Eule, S. and Metzger, J. J.},
  title = {Non-equilibrium steady states of stochastic processes with intermittent resetting},
  journal = {New J. Phys.},
  volume = {18},
  number = {3},
  pages = {033006},
  year = {2016},
  doi = {10.1088/1367-2630/18/3/033006},
  url = {https://iopscience.iop.org/article/10.1088/1367-2630/18/3/033006}
}

@article{Evans2018,
  author = {Evans, M. R. and Majumdar, S. N.},
  title = {Run and tumble particle under resetting: A renewal approach},
  journal = {J. Phys. A: Math. Theor.},
  volume = {51},
  number = {47},
  pages = {475003},
  year = {2018},
  doi = {10.1088/1751-8121/aae74e},
  url = {https://iopscience.iop.org/article/10.1088/1751-8121/aae74e}
}

@article{reuveni2016,
  title = {Optimal Stochastic Restart Renders Fluctuations in First Passage Times Universal},
  author = {Reuveni, Shlomi},
  journal = {Phys. Rev. Lett.},
  volume = {116},
  issue = {17},
  pages = {170601},
  numpages = {6},
  year = {2016},
  month = {Apr},
  publisher = {American Physical Society},
  doi = {10.1103/PhysRevLett.116.170601},
  url = {https://link.aps.org/doi/10.1103/PhysRevLett.116.170601}
}

@article{Pal_2016,
   title={Diffusion under time-dependent resetting},
   volume={49},
   ISSN={1751-8121},
   url={http://dx.doi.org/10.1088/1751-8113/49/22/225001},
   DOI={10.1088/1751-8113/49/22/225001},
   number={22},
   journal={J. Phys. A: Math. Theor.},
   publisher={IOP Publishing},
   author={Pal, Arnab and Kundu, Anupam and Evans, Martin R},
   year={2016},
   month=apr, pages={225001} }

@article{Masoliver2019,
  author = {Masoliver, J. and Montero, M.},
  title = {Anomalous diffusion under stochastic resetting: a general approach},
  journal = {Phys. Rev. E},
  volume = {100},
  number = {4},
  pages = {042103},
  year = {2019},
  doi = {10.1103/PhysRevE.100.042103},
  url = {https://link.aps.org/doi/10.1103/PhysRevE.100.042103}
}

@article{Bodrova2019,
  author = {Bodrova, A. S. and Chechkin, A. V. and Sokolov, I. M.},
  title = {Scaled Brownian motion with renewal resetting},
  journal = {Phys. Rev. E},
  volume = {100},
  number = {1},
  pages = {012120},
  year = {2019},
  doi = {10.1103/PhysRevE.100.012120},
  url = {https://link.aps.org/doi/10.1103/PhysRevE.100.012120}
}

@article{besga_2020,
  title = {Optimal mean first-passage time for a Brownian searcher subjected to resetting: Experimental and theoretical results},
  author = {Besga, Benjamin and Bovon, Alfred and Petrosyan, Artyom and Majumdar, Satya N. and Ciliberto, Sergio},
  journal = {Phys. Rev. Res.},
  volume = {2},
  issue = {3},
  pages = {032029},
  numpages = {5},
  year = {2020},
  month = {Jul},
  publisher = {American Physical Society},
  doi = {10.1103/PhysRevResearch.2.032029},
  url = {https://link.aps.org/doi/10.1103/PhysRevResearch.2.032029}
}

@Article{Tal_Friedman2020,
author={Tal-Friedman, Ofir
and Pal, Arnab
and Sekhon, Amandeep
and Reuveni, Shlomi
and Roichman, Yael},
title={Experimental Realization of Diffusion with Stochastic Resetting},
journal={J. Phys. Chem. Lett.},
year={2020},
month={Sep},
day={03},
publisher={American Chemical Society},
volume={11},
number={17},
pages={7350-7355},
doi={10.1021/acs.jpclett.0c02122},
url={https://doi.org/10.1021/acs.jpclett.0c02122}
}

@article{Faisant_2021,
doi = {10.1088/1742-5468/ac2cc7},
url = {https://dx.doi.org/10.1088/1742-5468/ac2cc7},
year = {2021},
month = {nov},
publisher = {IOP Publishing and SISSA},
volume = {2021},
number = {11},
pages = {113203},
author = {Faisant, F and Besga, B and Petrosyan, A and Ciliberto, S and Majumdar, Satya N},
title = {Optimal mean first-passage time of a Brownian searcher with resetting in one and two dimensions: experiments, theory and numerical tests},
journal = {J. Stat. Mech.: Theory Exp.}
}

@book{artin2011algebra,
  author    = {Michael Artin},
  title     = {Algebra},
  edition   = {2},
  publisher = {Pearson Education},
  address   = {Boston, MA},
  year      = {2011},
  isbn      = {978-0-13-241377-0},
  lccn      = {2010017573},
  oclc      = {607974243},
  url       = {https://archive.org/details/algebra00arti_226}
}

@article{VVS07,
	author = {Vergassola, Massimo and Villermaux, Emmanuel and Shraiman, Boris I.},
	date = {2007/01/01},
	doi = {10.1038/nature05464},
	id = {Vergassola2007},
	isbn = {1476-4687},
	journal = {Nature},
	number = {7126},
	pages = {406--409},
	title = { `Infotaxis' as a strategy for searching without gradients},
	url = {https://doi.org/10.1038/nature05464},
	volume = {445},
	year = {2007}}

@article{mesquita2025,
      title={Dynamically generated correlations in a trapped bosonic gas via frequency quenches}, 
      author={Nikhil Mesquita and Manas Kulkarni and Satya N. Majumdar and Sanjib Sabhapandit},
      year={2025},
      pages={2509.00487},
      journal={arXiv},
      url={https://arxiv.org/abs/2509.00487}, 
}

@article{delvecchio2025b,
      title={Proxitaxis: an adaptive search strategy based on proximity and stochastic resetting}, 
      author={{Del Vecchio Del Vecchio}, Giuseppe and Manas Kulkarni and Satya N. Majumdar and Sanjib Sabhapandit},
      year={2025},
      pages={2507.05800},
      journal={arXiv},
      primaryClass={cond-mat.stat-mech},
      url={https://arxiv.org/abs/2507.05800}, 
}

\end{document}